\documentclass[a4paper,fleqn,usenatbib]{mnras}
\usepackage[T1]{fontenc}
\usepackage{ae,aecompl}
\usepackage{amstext}
\usepackage{url}
\usepackage{graphicx}	
\usepackage{amsmath}	
\usepackage{amssymb}	

\newcommand{\bea}{\begin{eqnarray}}
\newcommand{\eea}{\end{eqnarray}}
\newcommand{\be}{\begin{equation}}
\newcommand{\ee}{\end{equation}}

\title[Magnetized Filament Plasma Lenses]{Magnetized Filament Models for Diverging Plasma Lenses}

\author[Rogers et al.]
	{Adam Rogers$^1$ \thanks{E-mail: rogers@physics.umanitoba.ca},
	Abdul Mohamed$^2$,
	Bailey Preston$^2$,
	Jason D. Fiege$^1$,
	Xinzhong Er$^3$ \thanks{E-mail: xer@ynu.edu.cn}
	\\
$^1$ Department of Physics and Astronomy, University of Manitoba, Winnipeg R3T 2N2, Canada\\
$^2$ Department of Physics and Astronomy, Brandon University, Brandon R7A 6A9, Canada\\
$^3$ South-Western Institute for Astronomy Research, Yunnan University, Kunming, P. R. China}

\date{Accepted 2020 February 03. Received 2020 February 01; in original form 2019 August 01}

\pubyear{2020}

\begin{document}
\label{firstpage}
\pagerange{\pageref{firstpage}--\pageref{lastpage}}
\maketitle

\begin{abstract}
Spherical plasma lens models are known to suffer from a severe over-pressure problem, with some observations requiring lenses with central pressures up to millions of times in excess of the ambient ISM. There are two ways that lens models can solve the over-pressure problem: a confinement mechanism exists to counter the internal pressure of the lens, or the lens has a unique geometry, such that the projected column-density appears large to an observer. This occurs with highly asymmetric models, such as edge-on sheets or filaments, with potentially low volume-density. In the first part of this work we investigate the ability of non-magnetized plasma filaments to mimic the magnification of sources seen behind spherical lenses and we extend a theorem from gravitational lens studies regarding this model degeneracy. We find that for plasma lenses, the theorem produces unphysical charge density distributions. In the second part of the work, we consider the plasma lens over-pressure problem. Using magnetohydrodynamics, we develop a non self-gravitating model filament confined by a helical magnetic field. We use toy models in the force-free limit to illustrate novel lensing properties. Generally, magnetized filaments may act as lenses in any orientation with respect to the observer, with the most high density events produced from filaments with axes near the line of sight. We focus on filaments that are perpendicular to the line of sight that show the toroidal magnetic field component may be observed via the lens rotation measure.

\end{abstract}


\section{Introduction}
\label{sec:intro}

Extreme scattering events (ESEs) are observed in radio sources that show smooth changes in flux density at radio wavelengths ($\approx 1$ GHz), often dimming by $> 50 \%$ and lasting from weeks to months. Despite decades of study, the ionized structures responsible for producing ESEs remain mysterious and elusive \citep{ISMReview}. It is generally believed that ESEs are produced by refraction from an electron density inhomogeneity along the line of sight that acts as a diverging lens \citep{ESE3}. Using light travel time arguments and a distance estimate to the lens, the light curve of the first observed ESE in the radio source QSO 0954+658 \citep{ESE0} implies an ionized lensing structure with radius $<7$ AU and central density $n_e \approx 4 \times 10^4$ cm$^{-3}$, assuming a spherical source.

The cornerstone for quantitative analysis of ESEs was placed by \citet[][CFL]{cleggFey} in developing the Gaussian plasma lens, a valuable model that is used to describe the optical behaviour of a discrete clump of ionized material in the ISM. The model still finds wide and novel applicability \citep{FRBplasma1}, and has been expanded on in a number of ways \citep{lens1, ESE2, erRogers, rogersEr, erRogers2}. The CFL lens is a spherically symmetric model that has a Gaussian radial electron density. The lens is capable of producing between one and three images of a source \citep{ESE1, mi2, mi3}. A wide morphology of light curve shapes is possible, provided an impact parameter is specified \citep{axisym, erRogers}. The model has been very successful at reproducing observed details of ESEs. In addition, some phenomena in the scintillation of pulsars may be related to refractive effects \citep{pa1, pa2, basu, psrESE, hill05}. Pulsar scintillation is observed at GHz frequencies and occurs on a time-scale of days with variability on the order of a few percent.

In general, there are two closely related substantial difficulties that must be addressed with isolated spherical lenses if they are to be a viable solution to the ESE phenomena. The main problem is the high volume-density of ionized charges at the center of a spherical plasma lens that is needed to decrease the flux to the observed levels. The lens densities in question span a huge range from unity up to $n_\text{e} \approx 10^5$ cm$^{-3}$ \citep{coles2015, cleggFey}. A collection of the volume-density, column-density and inferred radii of all known ESEs has been compiled by \citet{ISMReview}. The exact method for generating such small but extremely dense structures in the ISM is not known. This lens over-density problem leads to a significant complication. Let us approximate the ISM using the average temperature of typical ionized gas, on the order of $T \approx 10^4$ K. For lenses on the order of $1$ AU in radius we find the central pressure given by the ideal gas law,
\begin{equation}
\frac{P}{k_\text{B}}=nT,
\end{equation}
where $k_\text{B}$ is the Boltzmann constant. For such a lens, the central pressure must be on the order of $P_\text{lens}/k_\text{B} \approx 10^{9}$ K cm$^{-3}$. The warm ionized interstellar medium has $n_\text{H} \leq 0.4$ cm$^{-3}$. Given this density we estimate the pressure of the ISM as $P_\text{ISM}/k_\text{B} \approx 4000$ K cm$^{-3}$ \citep{heiles}. Thus, plasma lenses must be over-pressured by up to a million times compared to the ISM surrounding them. Such lenses either form in a high-pressure environment, require a confinement mechanism to prevent them from exploding, or are naturally transient on the order of years \citep{ISMReview}. Even the substructures responsible for the scintillation arclets of pulsars are victims of this over-pressure problem \citep{hill05}. Assuming spherical symmetry, both the over-density and over-pressure problems are ubiquitous among all ESEs observed to date.

The ongoing real-time ESE monitoring program ATESE, conducted by \citet{ESE3} using the Australia Telescope Compact Array (ATCA), has verified that ESEs seem to be the result of over-dense diverging plasma lenses. For example, an ESE was observed in the radio source PKS 1939-315 that requires an over-dense lens with $n_\text{e} \approx 10^3$ cm$^{-3}$, eliminating the possibility of ``exotic'' under-dense converging plasma lenses \citep{lens1} at least for this particular observation. Thus, it appears that despite the apparent theoretical difficulties with the standard interpretation of ESEs, nature must have some mechanism for efficiently confining plasma lenses.

An elegant and natural solution to the over-pressure problem was suggested by \citet{romani87} as the result of a projection effect. Lenses that are elongated along the line of sight can show high apparent column-density despite a relatively low volume-density, which reduces or outright eliminates the over-density problem in a clever way. Thus, both filaments and sheets have a long history \citep{goldESE} in plasma lens model building as an alternative to over-dense isolated spherical plasma lenses \citep{ESE2, spSheets, coles2015}.

Filamentary gravitational lenses were first investigated \citep{bozza05} to describe tiny clouds of cold self-gravitating gas that serve as a purely baryonic alternative explanation for dark matter \citep{p94, dp95}. The suggestion that primordial clouds were responsible for ESEs was developed by \citet{ww1998} and \citet{ww99} who suggested that a population of $\approx 10^{15}$ planetary-mass ($10^{-4} M_\odot$), AU-scale clouds could serve as a population of objects with the right properties for producing the observed refractive effects. If these neutral clouds were in hydrostatic equilibrium with thermal pressure and balanced by self-gravity they may be surrounded by an ionized enveloping wind, evaporated from the cloud by UV radiation. The free electrons in this ionized layer would provide frequency-dependent lensing of background radio sources. To avoid direct detection, these clouds would have to be cold, on the order of the temperature of the cosmic microwave background \citep{p94} and transparent to optical radiation. Constraints from gravitational microlensing on the properties of such clouds have been studied by \citet{hw95}, and \citet{bozza05} developed a more realistic filamentary gravitational lens model to account for the cylindrical structure of such putative clouds. Due to the difficulty in directly studying such ephemeral structures the material composition of these clouds is speculative \citep{w13} and they remain mysterious \citep{m01}.

\citet{romani87} suggested an association between ESEs and supernova remnants (SNRs). Old SNRs could provide a natural high pressure environment in which a high pressure compact lens object could survive by eliminating the radial pressure difference between the interior and exterior of the plasma lens. Additionally, the turbulent shells of old SNRs could provide a natural environment to produce sheets, filaments and knots, which are all useful as potential lens structures. Radio echoes of pulses from the Crab pulsar PSR B0531+21 have been observed from the filamentary structure of the Crab nebula \citep{gsCrab}, which relates refraction and the observed structures within SNR environs.

Extreme intra-day variability of radio sources has been suggested as a consequence of structures in the hot plasma surrounding nearby stars \citep{w17}. These structures have been posited to be similar to the cometary knots seen in planetary nebulae: dense, radially-oriented, elongated molecular clumps surrounded by a thin layer of ionized gas. Approximately one stellar mass worth of material is distributed among $\approx 10^{5}$ distinct structures to a distance of up to $1.75$ pc from the host star. In this picture, these molecular clumps are circumstellar features that are shaped by UV photons from the center star. The plasma is an ionized outflow from the dense neutral clumps that are gravitationally bound to the star. \citet{w17} have identified several radio sources which show strong intra-day variability that have stars well aligned with the line of sight. This relationship forms a connection between structures around hot stars and variability of background radio sources.

Our work is intended to complement the ``noodle model'' of pulsar scintillation arcs proposed by \citet{noodles} and elaborated on by \citet{noodles2}. The noodle model describes a lens comprised of thin strips in projection. The model provides the full lensing description in the wave optics limit and produces pulsar scintillation arcs through Kirchoff diffraction. In contrast, our work is based in the geometrical optics limit, appropriate for refractive lensing, and focuses on filaments as the refractive plasma lenses responsible for ESEs. To overcome the over-pressure problem, we study filamentary lens models that are confined by magnetic fields, though the filaments may also be tilted toward the observer. In Section \ref{sec:theory} we develop the background of plasma lensing and in Section \ref{sec:geometry}, we compare the differences between spherical and cylindrical plasma lens models. In Section \ref{sec:degen}, we will use the theory developed to extend a theorem originally presented by \citet{bozza05} which shows that a source passing behind a cylindrical gravitational lens can perfectly mimic the passage of an identical source behind a spherical gravitational lens if the correct cylindrical lens density is used. We will investigate the applicability of this theorem to plasma lenses and examine the physicality of the solutions produced. In Section \ref{sec:mag1} we follow \citet{noodles} and \citet{filamentRM} by including the effect of a magnetic field in our lens model. However, we generalize previous results to describe a filament confined by a helical magnetic field. Section \ref{sec:jason} describes a non self-gravitating model filament using magnetohydrodynamics (MHD) and includes an investigation of the filament lifetimes. We develop toy models in Section \ref{sec:models} to give an example of the lensing effects that can occur from magnetically confined filaments. Finally, we provide further discussion and summarize our conclusions in Section \ref{sec:discussionConclusions}.

\section{Theory}
\label{sec:theory}
Here we will discuss the background of plasma lensing, which is closely related to the mathematics of gravitational lensing. For the remainder of this work we will use CGS units, following the plasma lens literature. After we discuss basic lens theory we will specialize to spherical and cylindrical Gaussian plasma lenses in Section \ref{sec:geometry}. To begin, the phase difference caused by the lensing of rays through plasma is
\be
\Phi = \frac{2 \pi}{\lambda} \int_{-\infty}^\infty (n-1) dz
\label{potential}
\ee
where $\lambda$ is the wavelength we are observing at. We consider the optical medium to be described by the index of refraction
\be
n^2 = 1-\frac{\omega_\text{e}^2}{\omega^2} .
\ee
In the case of a cold, non-magnetized plasma we have the plasma frequency
\be
\omega_\text{e}^2 = \frac{4 \pi e^2 n_\text{e}(r)}{m_\text{e}}
\ee
where $n_\text{e}(r)$ is the volume-density of charges. In the limit that $\omega$ is large compared to $\omega_\text{e}$,
\be
\Phi = - \frac{2 \pi}{\lambda} \frac{1}{2\omega^2} \int_{-\infty}^\infty \omega_\text{e}^2 dz
\ee
using $\omega = 2\pi c / \lambda$ and the classical electron radius $r_\text{e} = e^2/m_\text{e} c^2$ we simplify to find
\be
\Phi = - \lambda r_\text{e} \int_{-\infty}^\infty n_\text{e}(r) dz.
\ee
Finally, we define the projected electron density along the line of sight (also known as dispersion measure, DM)
\be
N_\text{e}(\boldsymbol{\theta}) = \int_{-\infty}^\infty n_\text{e}(r) dz
\label{etaDef}
\ee
where we have projected the electron number volume-density onto the lens plane and expressed the projected electron density (DM) $N_\text{e}$ in the angular coordinates that describe the image positions on the plane of the sky, $\boldsymbol{\theta}=(\theta_x,\theta_y)$. With this definition, we have
\be
\Phi(\boldsymbol{\theta}) = - \lambda r_\text{e} N_\text{e}(\boldsymbol{\theta}).
\ee
Refractive lensing is described in the geometric optics limit using the reduced deflection angle
\be
\boldsymbol{\alpha} = \boldsymbol{\nabla}_\theta \Psi
\label{defl}
\ee
with the effective potential for plasma lenses in analogy with gravitational lensing \citep{ESE4, erRogers}
\be
\Psi(\boldsymbol{\theta}) =  \frac{\lambda^2}{2 \pi}\frac{D_{ds}}{D_s D_d} r_\text{e} N_\text{e}(\boldsymbol{\theta})
\ee
where $D_d$ and $D_s$ are the distances from the observer to the lens and the observer to the source, respectively. The distance between the lens and source is $D_\text{ds}$. With the lens potential, the thin lens equation is given by
\be
\boldsymbol{\beta}=  \boldsymbol{\theta}  - \boldsymbol{\alpha},
\ee
where $\boldsymbol{\beta}=(\beta_x,\beta_y)$ are the coordinates that describe the position of the source.

In the following discussion we will neglect the mass of the filamentary structure. In the first half of the work we will consider the refraction occurring entirely due to cold non-magnetized plasma. The strength of the lensing effect is set by the constant characteristic angular scale $\theta_0$, analogous to the angular Einstein radius in gravitational lensing. In general, the spherical and cylindrical cases (measured from the center of the filament) may have different scale angles, so throughout the text we will denote the spherical characteristic angular scale as $A_s$ and the cylindrical characteristic angular scale as $A_c$. In general, we use Cartesian components on the lens plane, and denote the radial distance from the origin as
\begin{equation}
    \theta=\sqrt{ \theta_x^2 + \theta_y^2 }.
\end{equation}
Let the cylinder be oriented such that the long axis runs along the $\theta_y$ direction on the plane of the sky. In this case, the deflection along the filament axis vanishes due to the symmetry of a very long filament. Such ``extremely anisotropic models'' were discussed by \citet{ESE4}, in which case $N_\text{e}=N_\text{e}(\theta_x)$. Then
\be
\frac{\partial \Psi}{\partial \theta_y}=0
\ee
and the thin lens equation is reduced to a one-dimensional deflection in the $\theta_x$ direction,
\be
\beta_x = \theta_x - \alpha(\theta_x) =  \theta_x - \frac{\text{d}\Psi}{\text{d}\theta_x}
\ee
\be
\beta_y = \theta_y.
\ee
The inverse magnification provided by a lens is given in terms of the Jacobian determinant of the thin lens equation. In this anisotropic case the magnification simplifies, such that the elements of the Jacobian become
\be
J=\left[
 \begin{array}{cc}
	1-\frac{\text{d}\alpha}{\text{d}\theta_x} & 0  \\
	0 & 1
 \end{array}
\right] ,
\label{Jtensor}
\ee
which produces the magnification
\be
\mu = \frac{1}{\text{det}(J)} = \frac{1}{1-\frac{\text{d}\alpha}{\text{d}\theta_x}}.
\ee
The elements of the Jacobian can be expressed in terms of convergence and shear. These quantities depend on the effective lensing potential,
\be
\kappa(\theta_x) = \frac{1}{2} \nabla^2 \Psi(\theta_x)
\label{eq:kappa}
\ee
and
\be
\gamma_1 = \frac{1}{2} \left( \frac{\partial^2 \Psi}{\partial \theta_x^2} -  \frac{\partial^2 \Psi}{\partial \theta_y^2} \right)
\ee
\be
\gamma_2 = \frac{\partial^2 \Psi}{\partial \theta_x \partial \theta_y}
\ee
with $\gamma = \sqrt{\gamma_1^2+\gamma_2^2}$ the shear magnitude. For a filament lens, we have
\be
\kappa(\theta_x) = \gamma_1=\frac{1}{2} \frac{\text{d} \alpha}{\text{d}\theta_x}
\ee
with $\gamma_2=0$.

\section{The Effect of Geometry}
\label{sec:geometry}

The change from spherical to cylindrical geometry introduces far-reaching consequences in the imaging and magnification behaviour of filament lenses.
In this section we will compare and contrast the differences between the lensing effects of spherical and cylindrical lenses that have Gaussian density profiles.
In order to compare these two distinct lens geometries fairly, we will assume both the spherical and cylindrical lens systems are configured with identical system geometry such that the distance between source and lens $D_{ds}$, the distance between observer and lens $D_d$, and the distance between the observer and source $D_s$ are the same for both lenses. Moreover, we also assume that both lenses are being observed at the same wavelength $\lambda$. Then the only difference between the cylindrical and spherical critical lens radii $A_c$ and $A_s$ are due to the analogously defined peak projected electron densities,
\begin{equation}
    A_i^2 = \lambda^2 \frac{1}{2\pi} \frac{D_{ds}}{D_d D_s} r_e N_{0i}
\end{equation}
where $N_{0s}$ and $N_{0c}$ are the maximum projected electron density of the spherical and the filament lenses, respectively.

As discussed at length in \citet{noodles}, a filament with moderate charge volume-density that is slanted toward the line of sight can produce a substantial projected charge column-density on the lens plane. In this configuration, the effective column density to an observer appears to change due to the projection of the cylinder lens,
\begin{equation}
    \int_{-\infty}^{\infty} n_\text{e} dz \rightarrow \int_{-\infty}^{\infty} n_\text{e} \csc(i) dz
\end{equation}
where $i$ is the angle of inclination between the long axis of the filament and the line of sight, normal to lens plane. Projection does not change the properties of the cylinder lens except for the apparent charge column-density within it. This provides an elegant, straightforward solution to the lens over-density and over-pressure problems \citep{romani87}.

\subsection{Spherical Gaussian Plasma Lens}
\label{sub:sphere}
Spherical lenses are naturally described in polar coordinates on the plane of the sky. For the spherical lens, we use the radial angular distance  $\theta=\sqrt{\theta_x^2+\theta_y^2}$ on the lens plane and $\beta$ on the source plane. Poisson's equation gives a relationship between the lens potential and convergence,
\begin{equation}
    \nabla^2 \Psi(\theta) = 2 \kappa(\theta).
\end{equation}
Making use of the connection between the deflection angle and lens potential, we get
\begin{equation}
    \nabla \cdot \mathbf{\alpha} = 2 \kappa(\theta),
\end{equation}
and with the definition of the divergence in polar coordinates we write the radial component as
\begin{equation}
\nabla \cdot \mathbf{\alpha} = \frac{1}{\theta} \frac{\partial }{\partial \theta}\left( \theta \alpha \right)
\end{equation}

Generally, lenses with spherical symmetry have a radial deflection angle that can be written in terms of the spherical convergence $\kappa_s$,
\begin{equation}
\alpha=\frac{2}{\theta} \int_0^\theta \kappa_s(\theta') \theta' d\theta'.
\end{equation}
A general spherical lens produces a magnification given by
\begin{equation}
    \mu_s^{-1}=\frac{\beta}{\theta} \left( \frac{d \beta}{d \theta} \right).
    \label{spMag}
\end{equation}

To use the expressions developed above we must choose a particular form for the projected electron density $N_\text{e}$ (the lens DM) across the plane of the sky. For a general exponential spherical lens, we choose
\be
N_\text{e}(\theta)=N_{0s} \exp\left(-\frac{\theta^h}{h\sigma^h} \right)
\ee
where $N_{0s}$ is the central density of this spherical lens, $h$ the power-index and $\sigma$ the width of the exponential lens. From this, we have the effective lens potential
\be
\psi=A_s^2 \exp\left(-\frac{\theta^h}{h \sigma^h}\right).
\ee
From this, all the other properties of the general exponential lens can be found including the spherical convergence,
\begin{equation}
    \kappa_s(\theta) = A_s^2 \left( \frac{\theta^h}{\sigma^h}-h \right)\frac{\theta^{h-2}}{2\sigma^h}\exp\left(-\frac{\theta^h}{h \sigma^h}\right)
\end{equation}
and shear
\begin{equation}
    \gamma_s(\theta) = \left(h-2-\frac{\theta^2}{\sigma^2} \right)A_s^2\frac{\theta^{h-2}}{2\sigma^h}\exp\left(-\frac{\theta^h}{h \sigma^h}\right)
    \label{sphericalGamma}
\end{equation}
which agrees with eqs. 49 and 50 in \citet{erRogers}. In this expression, the power-index $h$ determines how fast the lens convergence changes as a function of radial distance from the lens centre. Such an exponential lens has inverse magnification
\begin{equation}
\begin{array}{ll}
\mu^{-1} = & 1 + \dfrac{hA_s^2}{\sigma^h} \theta^{h-2} \left( 1-\dfrac{\theta^h}{h \sigma^h} \right) e^{ -\frac{\theta^h}{h\sigma^h}} \\
   &  + \dfrac{A_s^4}{\sigma^{2h}} \theta^{2(h-2)} \left( h - 1 - \dfrac{\theta^h}{\sigma^h} \right) e^{ - 2 \frac{\theta^h}{h \sigma^h}}.
\end{array}
\end{equation}

The spherical Gaussian plasma lens has $h=2$ where
\begin{equation}
    \alpha(\theta)=-A_s^2 \frac{\theta}{\sigma^2} \exp\left(-\frac{\theta^2}{2\sigma^2}\right)
\end{equation}
with convergence
\begin{equation}
    \kappa_s(\theta) = A_s^2 \left( \frac{\theta^2}{\sigma^2}-2 \right)\frac{1}{2\sigma^2}\exp\left(-\frac{\theta^2}{2 \sigma^2}\right) = A_s^2 \Sigma_s(\theta)
    \label{sphericalKappa}
\end{equation}
where we have used $\Sigma_s(\theta)$ as the scaled spherical convergence. The shear is given by
\begin{equation}
    \gamma_s(\theta) = -\left( \frac{\theta^2}{\sigma^2} \right)\frac{A_s^2}{2\sigma^2}\exp\left(-\frac{\theta^2}{2 \sigma^2}\right)
    \label{sphericalGamma}
\end{equation}
giving the Gaussian plasma lens magnification,
\begin{equation}
\begin{array}{ll}
\mu_s^{-1} = & 1 + \dfrac{2 A_s^2}{\sigma^2} \left( 1-\dfrac{\theta^2}{2 \sigma^2} \right) e^{ -\frac{\theta^2}{2\sigma^2}} \\
   &  + \dfrac{A_s^4}{\sigma^{4}} \left( 1 - \dfrac{\theta^2}{\sigma^2} \right) e^{ - \frac{\theta^2}{ \sigma^2}}.
\end{array}
\label{mus}
\end{equation}
The imaging properties of the spherical Gaussian lens depend on the characteristic scale angle of the lens $A_s$ and the width of the lens $\sigma$. In fact, it can be shown that there exists a critical lens width \citep{erRogers}, such that
\begin{equation}
\sigma_\text{crit}=\frac{A_s}{\sqrt{ \frac{\exp(3/2)}{2}}}.
\end{equation}
Sub-critical Gaussian lenses that have $\sigma<\sigma_\text{crit}$ do not produce any caustics, so there is no multiple imaging. As the width increases above this critical limit, two caustics appear, and the super-critical Gaussian lens produces three images. An example of multiple imaging from a super-critical spherical Gaussian lens is shown in the top panels of Figure \ref{fig:sphCylImg} as thick gray lines. The bottom row shows the corresponding position of the circular source. In this work we exclusively consider point sources, but the circular source contours show the effect of the lens very clearly.

\subsection{Cylindrical Gaussian Plasma Lens}
\label{sub:cyl}

An extended filament lens produces substantial differences from a spherical lens due the lens geometry. The filament itself is a long cylinder, such that the projected profile defines a stripe on the lens plane. Due to symmetry it is most convenient to use Cartesian coordinates on the lens plane, with the axis of the cylinder lens in the $\theta_y$ direction. Therefore, the deflection produced by the lens is one-dimensional, and we need only consider the $\theta_x$ position on the lens plane, and $\beta_x$ on the source plane. Since the $\theta_y$ coordinate is irrelevant for this lens geometry we will neglect it altogether when dealing with cylindrical lenses and set $\beta_y=\theta_y=0$ for convenience. In Cartesian coordinates, we have
\begin{equation}
    \nabla \cdot \mathbf{\alpha} =  \frac{\partial \alpha_x }{\partial \theta_x}.
    \label{poisCyl}
\end{equation}
Giving the filament convergence $\kappa_c = A_c^2 \Sigma_c$, where $\sigma_c(\theta_x)$ is the scaled convergence of the cylindrical filament lens. Here, $A_c$ sets the angular scale of the lens. Note the similarity with the gravitational lens expression in terms of the linear mass density $\lambda=M/L$, where $L$ is the length of the filament, as used in \citet{bozza05}. Equation \ref{poisCyl} gives the deflection angle of the cylinder lens,
\begin{equation}
    \alpha(\theta_x)=2 A_c^2 \int_0^{\theta_x} \Sigma_c(\theta_x') d \theta_x' .
    \label{deflFromSigma}
\end{equation}

\begin{center}
\begin{figure*}
\includegraphics[viewport=31 92 385 210, clip=true, scale=1.40]{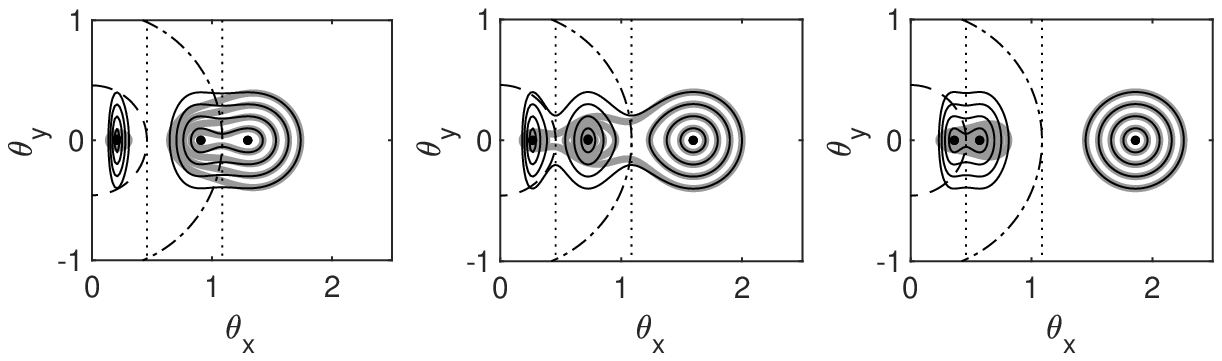}
\includegraphics[viewport=31 92 385 210, clip=true, scale=1.40]{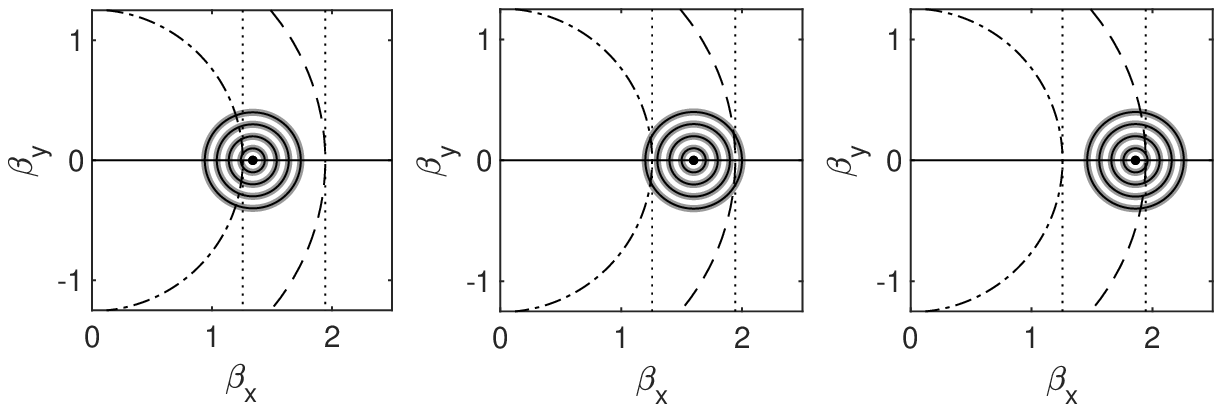}
\caption{Two examples of a super-critical Gaussian lens in spherical and cylindrical geometries. The source plane is shown in the panels on the bottom row. The lenses each act on a source comprised of a center point, and surrounded by concentric circles to emphasize the action of the lens. Both lenses produce two caustics on the source plane, shown as dashed and dash-dotted lines for the spherical lens and dotted lines for the cylindrical lens. The source is placed at $\beta_x=1.34$, $1.60$, $1.86$ from left column to right column, respectively. The source contours are shown in thick gray and thin black lines. The corresponding images and critical curves are shown in the image plane, on the top row of panels. The image contours are shown as thick gray lines (spherical lens) and thin black lines (cylindrical lens). We have chosen lenses with $A_c=A_s=1$ in arbitrary units to emphasize the difference geometry makes on producing lensed images. Both lenses have width $\sigma=1/2.5$.}
\label{fig:sphCylImg}
\end{figure*}
\end{center}

As an example, the cylindrical exponential lens gives the deflection angle
\begin{equation}
    \alpha(\theta_x) = -A_c^2\frac{\theta_x^{h-1}}{\sigma^h}\exp{\left(-\frac{\theta_x^h}{h \sigma^h} \right)}.
\label{eqCylDefl}
\end{equation}
The scaled convergence is
\begin{equation}
    \Sigma_c = \left(\frac{\theta_x^h}{\sigma^h}-h+1 \right)\frac{\theta_x^{h-2}}{2 \sigma^h}\exp{\left(-\frac{\theta_x^h}{h \sigma^h} \right)},
\end{equation}
which contains an extra term compared to the spherical convergence of an exponential lens (eq. \ref{sphericalKappa}). In the gravitational lens case, the lens convergence is interpreted as the projection of the mass distribution on the plane of the sky (the lens plane) and is a positive definite quantity. However, plasma lenses are capable of diverging lens behaviour, which implies the convergence can have both positive (converging) and negative (diverging) regions, and thus the convergence can no longer be interpreted as the projected mass for gravitational lenses (electron density for plasma lenses). A filament lens with a cylindrical Gaussian density profile ($h=2$) gives the deflection angle
\begin{equation}
    \alpha(\theta_x)=-A_c^2 \frac{\theta_x}{\sigma^2} \exp\left(-\frac{\theta_x^2}{2\sigma^2}\right),
\end{equation}
with filament magnification given by
\begin{equation}
\mu_c^{-1}=\left( \frac{d\beta_x}{d \theta_x} \right)=1-2 A_c^2\sigma_c \left( \theta_x \right)
\label{eq:muc}
\end{equation}
and the convergence $\kappa_c(\theta_x)=A_c^2 \Sigma_c(\theta_x)$ with
\begin{equation}
    \Sigma_c = \left(\frac{\theta_x^2}{\sigma^2}-1 \right)\frac{1}{2 \sigma^2}\exp{\left(-\frac{\theta_x^2}{2 \sigma^2} \right)}.
\end{equation}
To demonstrate the imaging properties of a cylindrical Gaussian plasma lens, we show an example of multiple imaging from a super-critical Gaussian lens in fig. \ref{fig:sphCylImg}. The top panels of the figure show the effect of the cylinder lens on a circular source plotted in thin black lines. The corresponding position of the circular source is shown in the panels on the bottom row.

\section{Spherical-Cylindrical Lens Degeneracy}
\label{sec:degen}

Let us compare the refractive effects of the spherical and cylindrical lenses. We follow the basic approach of \citet{bozza05} and will not consider lensing events in which sources cross caustics in the source plane. For the remainder of this section we restrict ourselves to sub-critical Gaussian lenses. However, as in the gravitational lens case, the theorem can be extended to include multiple imaging. See the appendix of \citet{bozza05} for details.

In this section our analysis relies on specifying the source and image positions, as well as the magnification produced by both spherical and cylindrical lenses individually. To keep our notation clear, we label all quantities related to the cylindrical lens with a subscript $c$ and all quantities related to the spherical lens with the subscript $s$. For example in this notation, a radial angular distance on the lens plane is given by $\theta_s=\sqrt{\theta_{sx}^2 +\theta_{sy}^2}$, whereas the cylinder lens in Cartesian coordinates has simply $\theta_c=\theta_{cx}$ since the $\theta_{cy}$ coordinate is irrelevant for cylindrical lenses as discussed in Section \ref{sub:cyl}.

We seek a condition on our cylindrical lens for which
\begin{equation}
    \mu_c(\theta_c(\beta_c))=\mu_s(\theta_s(\beta_s)).
\end{equation}
Using the definitions of spherical and cylindrical magnification in eqs. \ref{spMag} and \ref{eq:muc} respectively, we will derive an expression for the convergence that gives our cylindrical lens a magnification that identically matches the magnification of a spherical lens. This establishes an important degeneracy between spherical and cylindrical gravitational lenses. However, as we shall see, the required cylindrical convergence may not correspond to a physically realistic plasma filament. Regardless, we extend this theorem from gravitational to plasma lenses here.

If we let
\begin{equation}
    A_c^2 \Sigma_c(\theta_c)=\frac{1}{2}\left( 1-\frac{1}{\mu_s(\theta_s(\theta_c))}\right),
    \label{eq:cylConv}
\end{equation}
then our cylindrical plasma filament can match a spherical plasma lens provided we could relate the position of an image formed behind a spherical lens in terms of the corresponding image position of a source behind the cylindrical lens, ie $\theta_s(\theta_c)$.

Let an object passing behind a spherical lens have velocity $v_s$ in the x-direction, such that the x-component of the trajectory is $\beta_{sx}=v_s t$ with the impact parameter $\beta_{sy}=b$. We define the distance of a source from the lens center on the source plane as
\begin{equation}
    \beta_s^2=b^2+v_s^2 t^2.
    \label{spherical_src}
\end{equation}
Now, let an identical object pass behind the cylindrical lens with speed $v_\text{c}$, giving $\beta_\text{cx}=v_\text{c} t$. There is no corresponding impact parameter for the cylindrical lens since all linear source trajectories that pass behind the cylinder lens cross the symmetry axis (analogous to passing through the origin of a spherical lens). Let us find the time $t$ in terms of the source position behind the cylindrical lens,
\begin{equation}
    t=\frac{\beta_\text{c}}{v_\text{c}}.
\end{equation}
We use this expression to eliminate $t$ in eq. \ref{spherical_src}, giving
\begin{equation}
    \beta_s^2=\beta_\text{sx}^2+\beta_\text{sy}^2=b^2+r_\text{v}^2 \beta_\text{c}^2.
\end{equation}
with $r_\text{v}=v_\text{s} / v_\text{c}$. Re-arranging we find
\begin{equation}
    \beta_\text{c}= \pm \frac{\left( \beta_\text{s}^2 - b^2\right)^\frac{1}{2}}{r_\text{v}}.
\end{equation}
Taking the derivative of this expression yields
\begin{equation}
    \frac{\text{d}\beta_\text{c}}{\text{d}\theta_\text{c}}=\pm \frac{\beta_\text{s}}{(\beta_\text{s}^2-b^2)^\frac{1}{2}}\frac{1}{r_\text{v}}\frac{\text{d}\theta_\text{s}}{\text{d} \theta_\text{c}}\frac{\text{d} \beta_\text{s}}{\text{d} \theta_\text{s}}.
\end{equation}
Using the equality of the spherical and cylindrical magnification to eliminate the derivatives gives
\begin{equation}
    \frac{\text{d}\beta_\text{c}}{\text{d}\theta_\text{c}}= \frac{\beta_\text{s}}{\theta_\text{s}}\frac{\text{d} \beta_\text{s}}{\text{d} \theta_\text{s}}
\end{equation}
and we are left with
\begin{equation}
    \frac{\text{d} \theta_\text{s}}{\text{d} \theta_\text{c}}= \pm \frac{r_\text{v}}{\theta_\text{s}} \left(\beta_\text{s}^2-b^2 \right)^\frac{1}{2}.
    \label{ODE}
\end{equation}
This expression relates the image positions from a spherical lens $\theta_s$ to those produced by a cylinder lens $\theta_c$. We use the solution of this differential equation, $\theta_s(\theta_c)$, to evaluate eq. \ref{eq:cylConv}.

\begin{center}
\begin{figure}
  \includegraphics[viewport=19 23 390 278, clip=true, scale=0.6]{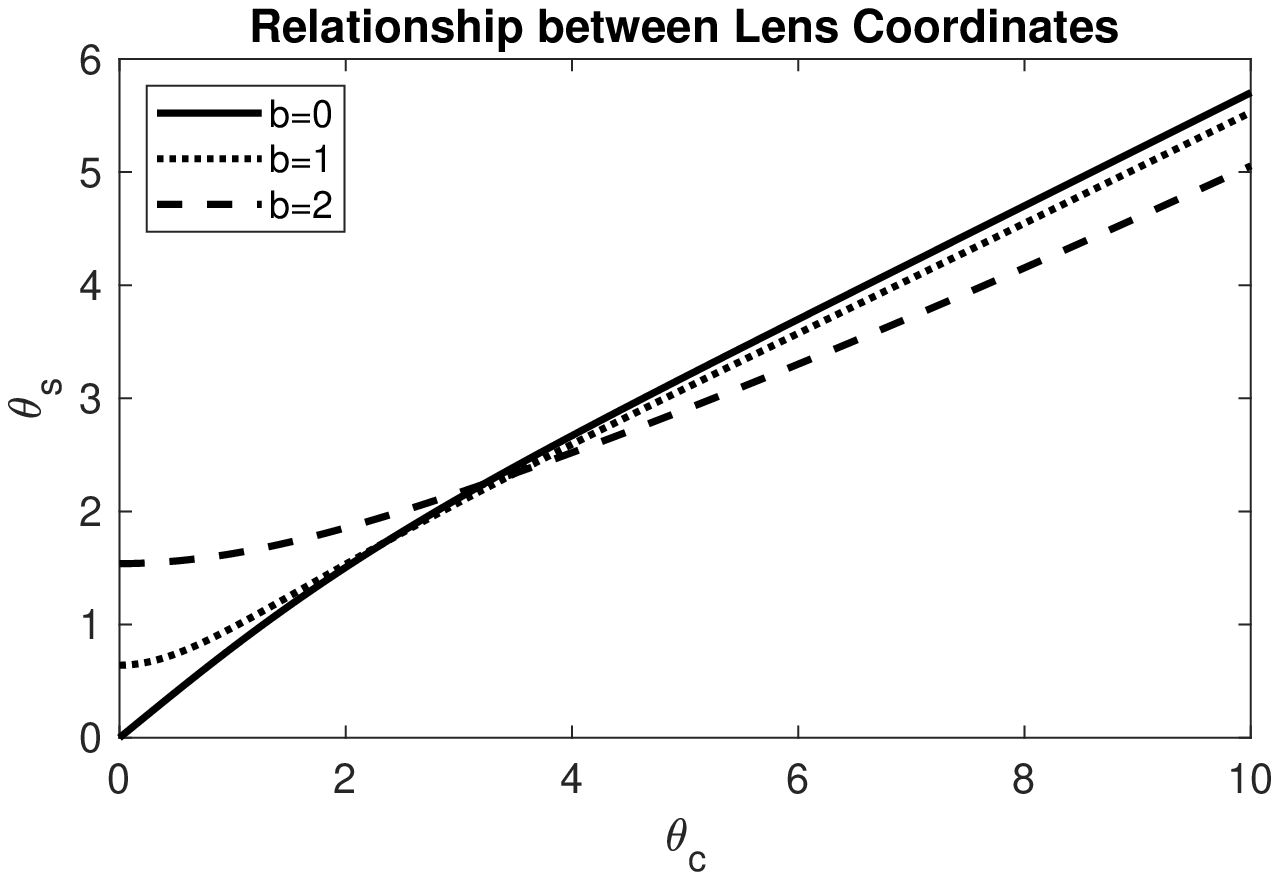}
  \includegraphics[viewport=50 0 350 308, clip=true, scale=0.65]{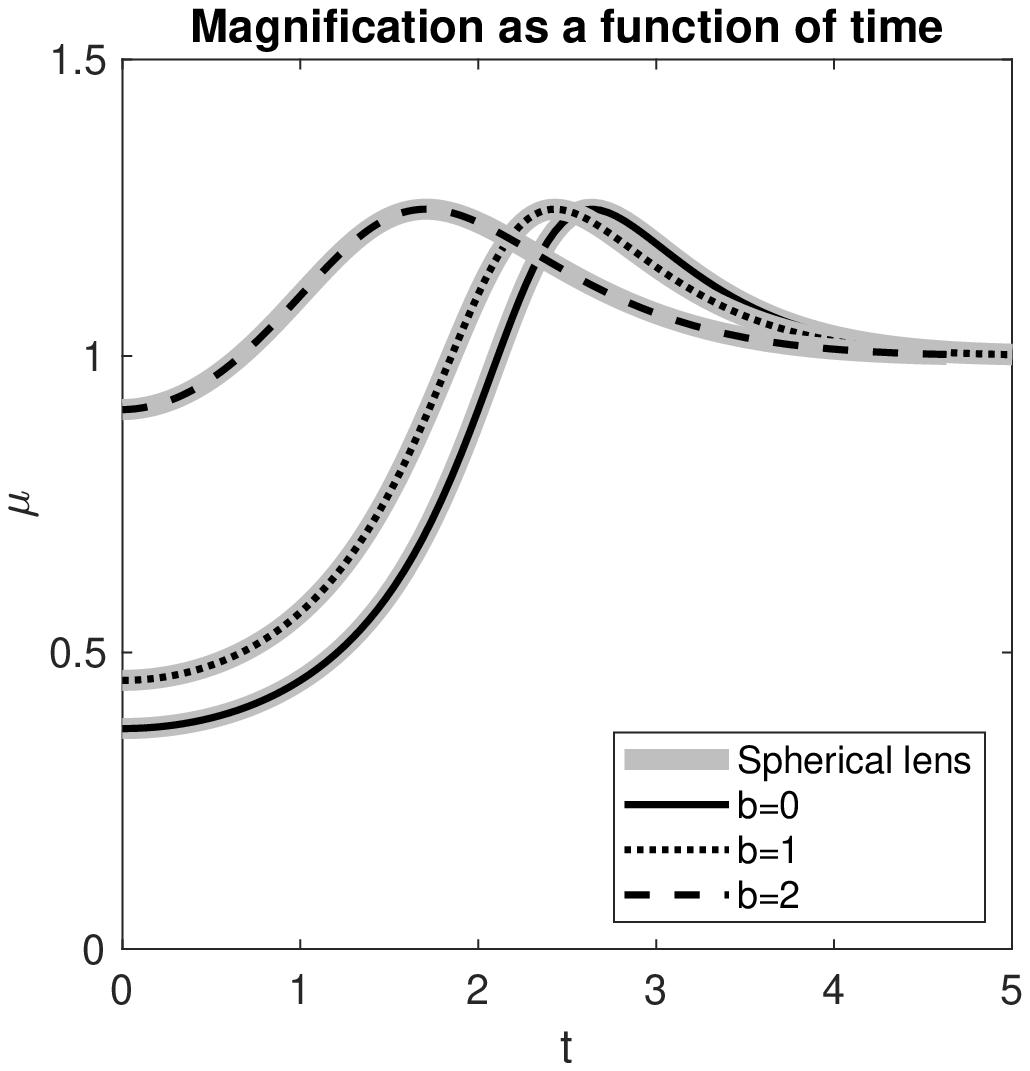}
  \includegraphics[viewport=50 0 350 308, clip=true, scale=0.65]{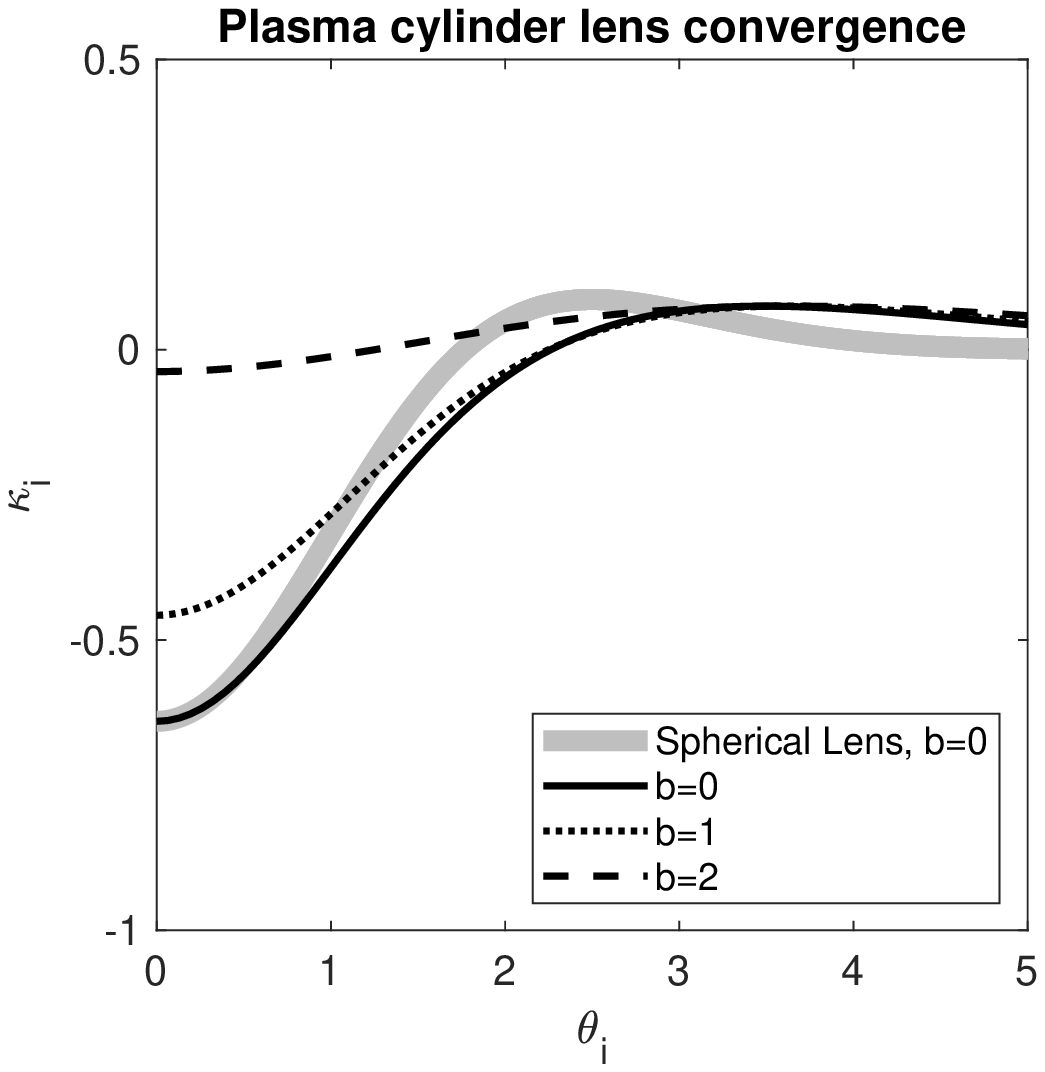}
  \caption{Top panel: The solution to the differential equation \ref{ODE} showing the relationship $\theta_s(\theta_c)$. We use a relative lens speed $r_v=1/2$ for impact parameters $b=0$ (solid line), $b=1$ (dotted line), $b=2$ (dashed line). We use this plotting scheme to distinguish the impact parameters in all three plots. Middle panel: The magnification in spherical $\mu_s$ (thick gray line) and cylindrical coordinates $\mu_c$ (black lines) as a function of time for the three impact parameters in the top panel. All curves are indistinguishable from one another. Bottom Panel: Cylindrical lens ($i=c$) convergence $\kappa_c$ plotted against $\theta_c$ assuming the three impact parameters (black lines) and spherical lens ($i=s$) convergence $\kappa_s$ (thick gray line) for $b=0$, plotted against $\theta_s(\theta_c)$. We have labelled the axes as $\kappa_i$ and $\theta_i$ since we plot both cylindrical and spherical quantities on this plot.}
\label{figCoords}
\end{figure}
\end{center}

The solution of the differential equation (eq. \ref{ODE}) is shown in the top panel of fig. \ref{figCoords} as a function of impact parameter, with $b=0$ (solid line), $b=1$ (dotted line) and $b=2$ (dashed line). The initial conditions for the solutions shown in the figure are the images of a source at a given impact parameter, $\theta_\text{s}(0)=\beta_\text{s}^{-1}(b)$. Even in the case of a sub-critical lens, the image position of the source is modified slightly from its vacuum position $\theta_\text{s}(0)=b$. The difference between the unlensed and lensed image positions decrease as the impact parameter of the source is increased. In the center panel of fig. \ref{figCoords} we compare the spherical and cylindrical magnifications $\mu_\text{c}$ and $\mu_\text{s}$ (though the two light curves are indistinguishable from one another), and in the lower panel the cylindrical convergence $\kappa_\text{c}$.

\begin{figure}
\begin{center}
  \includegraphics[viewport=46 4 350 305, clip=true, scale=0.75]{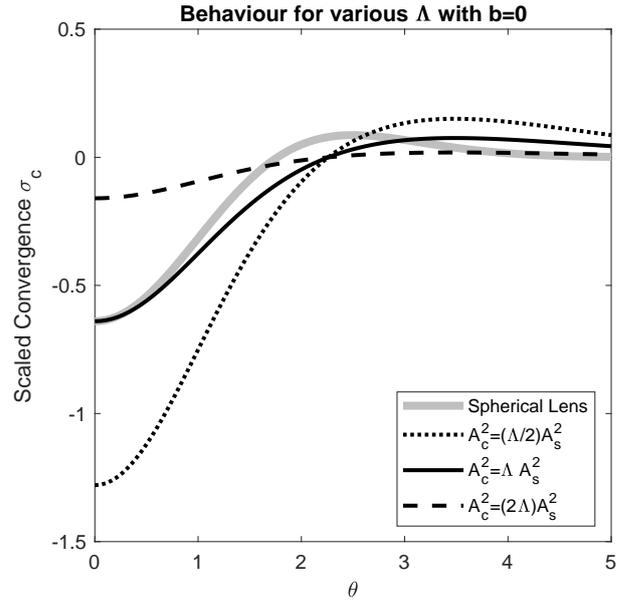}
  \caption{Critical behaviour of $\Lambda$. This plot shows the convergence of the spherical plasma lens as a solid gray line. The cylinder lens with $\Lambda<\Lambda_\text{crit}$ (dotted line) has a larger convergence (in magnitude) than the spherical lens. The critical-convergence cylinder lens has $\Lambda=\Lambda_\text{crit}$ (solid black line) and matches the spherical convergence at the center of the lens. The cylinder lens with $\Lambda>\Lambda_\text{crit}$ and has a lower central convergence than the spherical model.}
\label{fig:degen}
\end{center}
\end{figure}

Keeping in mind that we have restricted ourselves to non-caustic crossing lensing events, we can use the spherical lens magnification at the origin (impact parameter $b=0$) to get an estimate of the lensing convergence there.  For a sub-critical lens without any caustics, a source at the origin produces a demagnified image there. However, the plasma lensing case is somewhat more subtle than the gravitational lens case. First, we write the spherical magnification in terms of the convergence and shear \citep{narayan},
\begin{equation}
    \mu_s^{-1}(0)=(1-\kappa_s(0))^2-\gamma_s^2(0).
\end{equation}
with
$\gamma_s(0)=0$. Using the expression in eq. \ref{eq:cylConv}, we find the scaled filament convergence,
\begin{equation}
    \Sigma_\text{c}(0)=\frac{1}{2A_\text{c}^2}(1-(1-\kappa_\text{s}(0))^2)
\end{equation}
Simplifying this expression we find
\begin{equation}
    \Sigma_\text{c}(0)=\frac{\kappa_\text{s}(0)}{A_\text{c}^2}\left(1-\frac{\kappa_\text{s}(0)}{2}\right)
\end{equation}
Now, let us write the cylindrical scale length as some multiple of the spherical scale length, $A_\text{c}^2=\Lambda A_\text{s}^2$. These scales quantify the strength of the lenses and do not rely on the particular geometry. Therefore, we simplify the previous expression to find,
\begin{equation}
    \Sigma_\text{c}(0)=\frac{\Sigma_\text{s}(0)}{\Lambda}\left(1-\frac{A_\text{s}^2\Sigma_\text{s}(0)}{2}\right)
    \label{eq:sigCrit}
\end{equation}
which is totally general without any definition yet specified for $\Sigma_\text{c}(\theta_\text{c})$ and $\Sigma_\text{s}(\theta_\text{s})$. The above expression shows that there exists a critical ratio of the squares of scale angles between the spherical and cylindrical plasma lens models,
\begin{equation}
    \Lambda_\text{crit}= 1-\frac{A_s^2\Sigma_s(0)}{2}. 
\end{equation}
This critical value determines when the cylinder will have a larger central convergence compared to the spherical model. From eq. \ref{eq:sigCrit}, we see that when $\Lambda=\Lambda_\text{crit}$, the central convergence of spherical and cylindrical lens are equal to one another, $\Sigma_c(0)=\Sigma_s(0)$. Using the convergence for our particular choice of Gaussian lens, we find from eq. \ref{sphericalKappa} the spherical convergence at the center of the lens,
\begin{equation}
    \kappa_\text{s}(0)=-\frac{A_\text{s}^2}{\sigma^2},
\end{equation}
which gives the critical ratio
\begin{equation}
    \Lambda_\text{crit}=1+\frac{A_\text{s}^2}{2 \sigma^2}.
\end{equation}
Below this limit, $\Lambda<\Lambda_\text{crit}$, the cylindrical lens will have higher magnitude scaled convergence (ie, more negative) than a spherical model. When $\Lambda=\Lambda_\text{crit}$, the central convergence of both models are equal. Above the critical limit, $\Lambda>\Lambda_\text{crit}$, and the cylinder has a smaller in magnitude (ie, more positive) convergence than the spherical lens. These cases are demonstrated in fig. \ref{fig:degen}. Similar to the gravitational case studied by \citet{bozza05}, the required lensing convergence of the cylindrical lens decreases further for non-zero impact parameter.

\subsection{Physicality of Solutions}
\label{sec:problem}
\begin{figure}
\begin{center}
  \includegraphics[viewport=46 4 350 305, clip=true, scale=0.75]{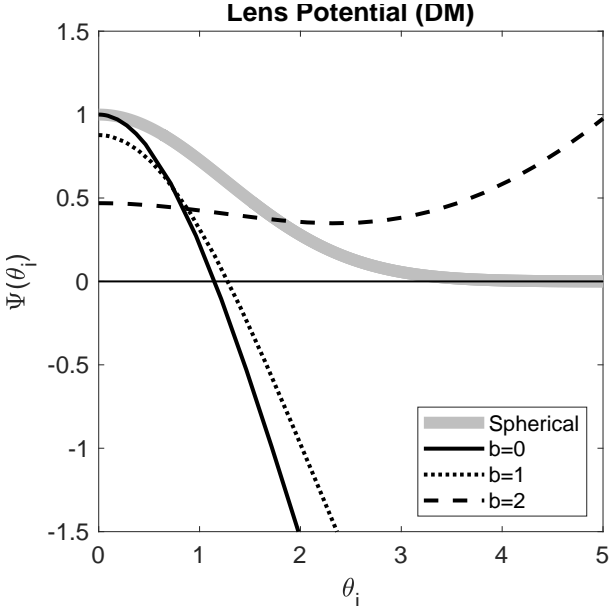}
  \caption{We show the lens potentials for the spherical Gaussian lens (thick gray line) plotted against $\theta_s(\theta_c)$, and the equivalent cylinder lens for a source with the impact parameters $b=0,1,2$ (black lines) plotted against $\theta_c$. We have labelled the axes with the subscript $i$ since we plot both cylindrical and spherical quantities on this plot. The horizontal thin black line is the axis, showing that the cylindrical solutions can be negative or growing with distance from the origin, both behaviours demonstrating unphysical characteristics.}
\label{fig:lensRecon}
\end{center}
\end{figure}

We have shown above that given a source passing behind a spherical plasma lens with any impact parameter, we can find a cylindrical projected charge density that will produce an identical magnification as the spherical lens. However, in practice we must examine the solutions that this theorem produces to ensure they are physically realistic. In fact, our numerical results show that the difference between plasma lenses and gravitational lenses are substantial in this regard. In the gravitational lens version of this theorem, \citet{bozza05} show that the filament gravitational lens always requires less central mass density than the equivalent spherical gravitational lens models. With gravitational lenses the convergence is interpreted as the projected lens mass density, which is always a positive quantity that produces converging lens magnification. With plasma lenses, both converging (magnifying) and diverging (demagnifying) lensing behaviour is seen, depending on the relative position of source, lens and observer. Since a plasma lens acts like a diverging lens at certain positions on the lens plane, the convergence at these positions must be negative. For plasma lenses, the convergence is not interpreted as the projected lens density as it is for gravitational lenses, instead it is the lens potential that depends on the projected charge density (DM). This is a significant difference between gravitational lenses and plasma lenses. Additionally, while negative densities are forbidden in gravitational lensing, negative DM is interpreted as an under-dense region of plasma, and allows for converging lensing to occur.
To find the effective lens potential we must use the deflection angle, which is the gradient of the lens potential, eq. \ref{defl}. For the cylinder lens we can write simply
\be
\Psi(\theta_c)=\int_0^{\theta_c} \alpha(\theta'_c) d \theta'_c
\ee
Since we find the deflection angle using eq. \ref{deflFromSigma} for our derived convergence $\Sigma_c$ from eq. \ref{eq:cylConv}, we can numerically integrate to find the cylindrical lens potential $\Psi(\theta_c)$, which depends directly on the lens DM. For all of the impact parameters we have tested, our cylindrical lens potential always contains features that are difficult to interpret physically. For example, the $b=0$ and $b=1$ solutions near the origin closely match the spherical Gaussian lens potential, 
\be
\Psi(\theta_s)=A_s^2 \exp \left(-\frac{\theta_s^2}{2\sigma^2} \right)
\ee
however the cylindrical density profile for these impact parameters plunge and quickly become negative. This represents an over-dense filament, with an under-dense surrounding environment that contributes an increasing converging effect, even far from the lens. As the impact parameter increases, the tail of this function also increases such that at $b=2$, we find a cylinder lens density that drops to a local minimum before rising with distance from the filament. The lens potentials for the spherical lens and the cylinders to reproduce $b=0$, $b=1$ and $b=2$ lenses are shown in fig. \ref{fig:lensRecon}. 

While the theorem does produce numerical solutions that demonstrate the spherical-cylindrical degeneracy for plasma lenses, the solutions are not physically plausible. Thus, we find it unlikely that physical filamentary plasma lenses oriented perpendicular to the line of sight could reproduce the magnification of spherical plasma lenses, despite the result for gravitational lenses. Hence, cylindrical plasma lenses viewed perpendicular to the axis should produce magnification patterns distinct from spherical lenses. Of course, there is a simple geometric way for a cylindrical lens to match a spherical lens, which is by projection. If a finite-length filament with circular cross-section is viewed end-on or at a low inclination with respect to the line of sight, it will appear similar to a dense sphere in projection.

\section{Magnetized Filaments as Plasma Lenses}
\label{sec:mag1}

To include a magnetic field in a plasma lens, we must use the magnetized index of refraction to determine the phase added to a wave due to passage through the lens,
\be
\Phi = \frac{2 \pi}{\lambda} \int_{-\infty}^\infty (n-1) dz
\ee
as in eq. \ref{potential}, except now we follow \citet{filamentRM} and make use of the index of refraction that includes a contribution from the magnetic field
\be
n^2 \approx 1-\left( \frac{\omega_\text{e}}{\omega} \right)^2 \left( \frac{1}{1 \pm \frac{\omega_\text{B}}{\omega}} \right)
\ee
where the electron cyclotron frequency is
\be
\omega_\text{B}=\frac{e|B|}{m_\text{e} c}.
\ee
The $+$ and $-$ signs correspond to right and left-hand circular polarizations respectively. Following \citet{filamentRM}, we identify the right-handed circularly polarized wave as having positive helicity.

The effective lens potential, in analogy with gravitational lensing, is
\be
\Psi =  -\frac{\lambda}{2 \pi}\frac{D_\text{ds}}{D_\text{s} D_\text{d}} \Phi
\ee
such that for the magnetized filament, the phase factor is
\be
\Phi=-\lambda r_\text{e} DM \mp \lambda^2 RM,
\ee
with the definitions of dispersion measure (DM)
\be
DM=\int_{-\infty}^{\infty} n_\text{e}(r) dz
\ee
and and rotation measure (RM)
\be
RM = \frac{r_\text{e}^2}{2\pi e} \int_{-\infty}^{\infty}n_\text{e} B_{||}(r,\phi,z) dz.
\label{RMdefn}
\ee
From here, we follow the usual derivation given in Section \ref{sec:theory} to arrive at the effective lens potential in analogy with gravitational lensing
\be
\Psi  =  \frac{\lambda^2}{2 \pi}\frac{D_\text{ds}}{D_\text{s} D_\text{d}} r_\text{e} DM \pm  \frac{\lambda^3}{2 \pi} \frac{D_\text{ds}}{D_\text{s} D_\text{d}} RM
\label{potDefn}
\ee
In terms of the lens plane coordinates, we write distance on the sky in terms of the angular coordinate $D_d \theta$ and $z$ the distance along the undeflected ray using the Born approximation for weak deflection. We calculate the deflection angle for a given $n_\text{e}$ and $\mathbf{B}$,
\be
\begin{array}{lll}
\alpha& = & \frac{\lambda^2}{2 \pi} \frac{D_\text{ds}}{D_\text{s} D_\text{d}} r_\text{e} \nabla_\theta \int_{-\infty}^{\infty} n_\text{e}(r) dz \\
  &  \pm & \lambda^3 \frac{D_\text{ds}}{D_\text{s} D_\text{d}} \frac{r_\text{e}^2 }{4 \pi^2 e} \nabla_\theta \int_{-\infty}^{\infty} \left[ n_\text{e}(r) B_{||}(r,\phi,z) \right] dz
  \label{deflGen}
\end{array}
\ee
A positive $B_{||}$ means the magnetic field is pointing toward the observer, while a field that points away is negative by convention. Since the filament lens is symmetric, the deflection angle along the axial direction vanishes and the lens remains one-dimensional. 

\subsection{Filaments with Poloidal Magnetic Fields}

The simplest case of a magnetized filament has a poloidal (axial) field $B_\text{z}$ that runs along the length of the filament in the direction $\hat{e}_\text{z}$. Such a case was considered in the noodle model for scintillation arcs \citep{noodles}. In that work, the magnetic field everywhere in the filament is orthogonal to the path of a passing ray, such that $B_{||}=0$, and there will be no rotation measure that can be observed from the filament. In terms of refractive lens phenomena, there will be no observable difference between the poloidal magnetic field and the non-magnetized filaments. Examples of field aligned HI fibers in the cold neutral medium \citep{ISMReview} have been discovered in the diffuse ISM \citep{h1fil}. These linear features vary in scale up to a maximum length $\sim10^4$ AU and are aligned along the poloidal magnetic field lines \citep{h2fil}. If the HI fibers can act as refractive lenses, they are most similar to the poloidal magnetic filament model \citep{noodles}.

\subsection{Magnetically Skewered Filaments}

A relatively straightforward magnetic field geometry is a skewered filament, in which magnetic field lines pierce the filament perpendicular to its symmetry axis. This is the case studied by \citet{filamentRM}, who take the magnetic field to have a constant component in the direction of ray propagation $B_{||}=B_0 \hat{z}$ along the line of sight. Using this component of the  field the deflection angle becomes
\begin{equation}
\alpha =  \left( 1
\pm  \lambda \frac{r_\text{e} }{2 \pi e} B_0 \right) \frac{\lambda^2}{2 \pi} \frac{D_\text{ds}}{D_\text{s} D_\text{d}} r_\text{e} \nabla_\theta \int_{-\infty}^{\infty}  n_\text{e}(r) dz.
\label{deflSkew2}
\end{equation}
Comparing with the general expression for the cylindrical lens deflection angle (eq. \ref{deflGen}), we write the deflection angle due to the DM term as $\alpha_\text{DM}(\theta_\text{x})$ from eq. \ref{deflGen}, this simplifies to
\begin{equation}
\alpha = \left( 1 \pm  \lambda \frac{r_e }{2 \pi e} B_0 \right) \alpha_\text{DM}(\theta_\text{x})
\label{deflSkew2}
\end{equation}
and we see that the magnetic field modifies the strength of the lens deflection for each polarization state independently, but otherwise only affects the refractive properties of the lens by changing the overall magnitude of the deflection (amounting to a redefinition of the characteristic scale angle $\theta_0$).

\subsection{Filaments with Helical Magnetic Fields}

We now consider an ionized, magnetized filament, confined by a helical magnetic field. To describe helical fields around filamentary lenses, we introduce a new cylindrical coordinate system that is centered on the lens itself. With this set of coordinates we define radial distances $r$ in the $\hat{e}_r$ direction, the azimuthal $\hat{e}_\phi$ direction and the $\hat{e}_z$ direction coaxial to the filament and perpendicular to the line of sight for simplicity. It is with respect to these cylindrical coordinates that we define the filament magnetic field. Just as in the non-magnetized case, the projection of the filament on the lens plane produces a strip that we describe with Cartesian coordinates.

\begin{figure}
\begin{center}
  \includegraphics[width=13.0 cm]{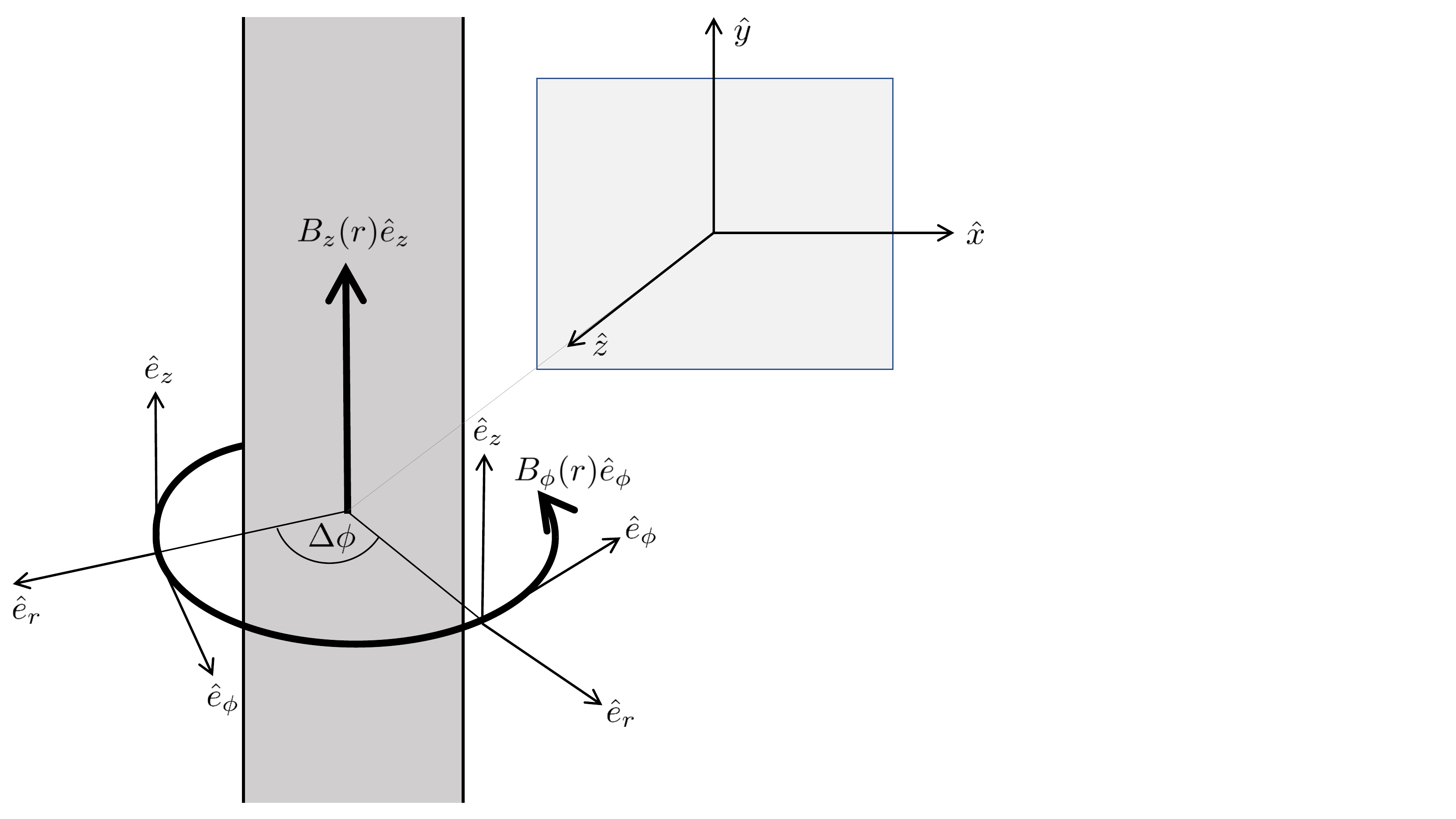}
  \caption{An illustration of the lensing geometry for a magnetized filament. The filament has both poloidal $B_\text{z}(r)$ and toroidal $B_\phi(r)$ components. The unit vector $\hat{z}$ points along the line of sight from the source	plane toward the observer. The cylinder is oriented such that the poloidal field along the cylinder axis $\hat{e}_\text{z}$ is parallel with the $\hat{y}$ unit vector on the source and image planes. The cylindrical polar unit vectors $\hat{e}_\text{r}$ and $\hat{e}_\phi$ change orientation as we change the equatorial angle $\Delta \phi$ and revolve around the filament. }
\label{figGeom2}
\end{center}
\end{figure}

Let the filament helical magnetic field be comprised of two components, an axial component $B_\text{z}$ that runs along the length of the filament in the direction of $\hat{e}_\text{z}$, and  a toroidal component $B_\phi$ in the $\hat{e}_\phi$ direction which determines the field in the plane perpendicular to the long axis of the filament:
\be
\mathbf{B} = B_\phi(r) \hat{e}_\phi + B_\text{z}(r) \hat{e}_\text{z}.
\ee
In fig. \ref{figGeom2} we illustrate the magnetic field components of the filament with the cylindrical coordinate system of the lens, as well as the relationship with the coordinates of the source plane. In general, we use the Cartesian unit vectors to describe directions on the source and image planes, with $\hat{x}$ and $\hat{y}$ such that $\boldsymbol{\theta}=\theta_x \hat{x} + \theta_y \hat{y}$ and $\boldsymbol{\beta}=\beta_x \hat{x} + \beta_y \hat{y}$. We use the same orientation of unit vectors to describe both image and source coordinates for convenience. The $\hat{z}$ vector points along the line of sight from source to observer.

\begin{figure}
\begin{center}
\includegraphics[width=13.0 cm]{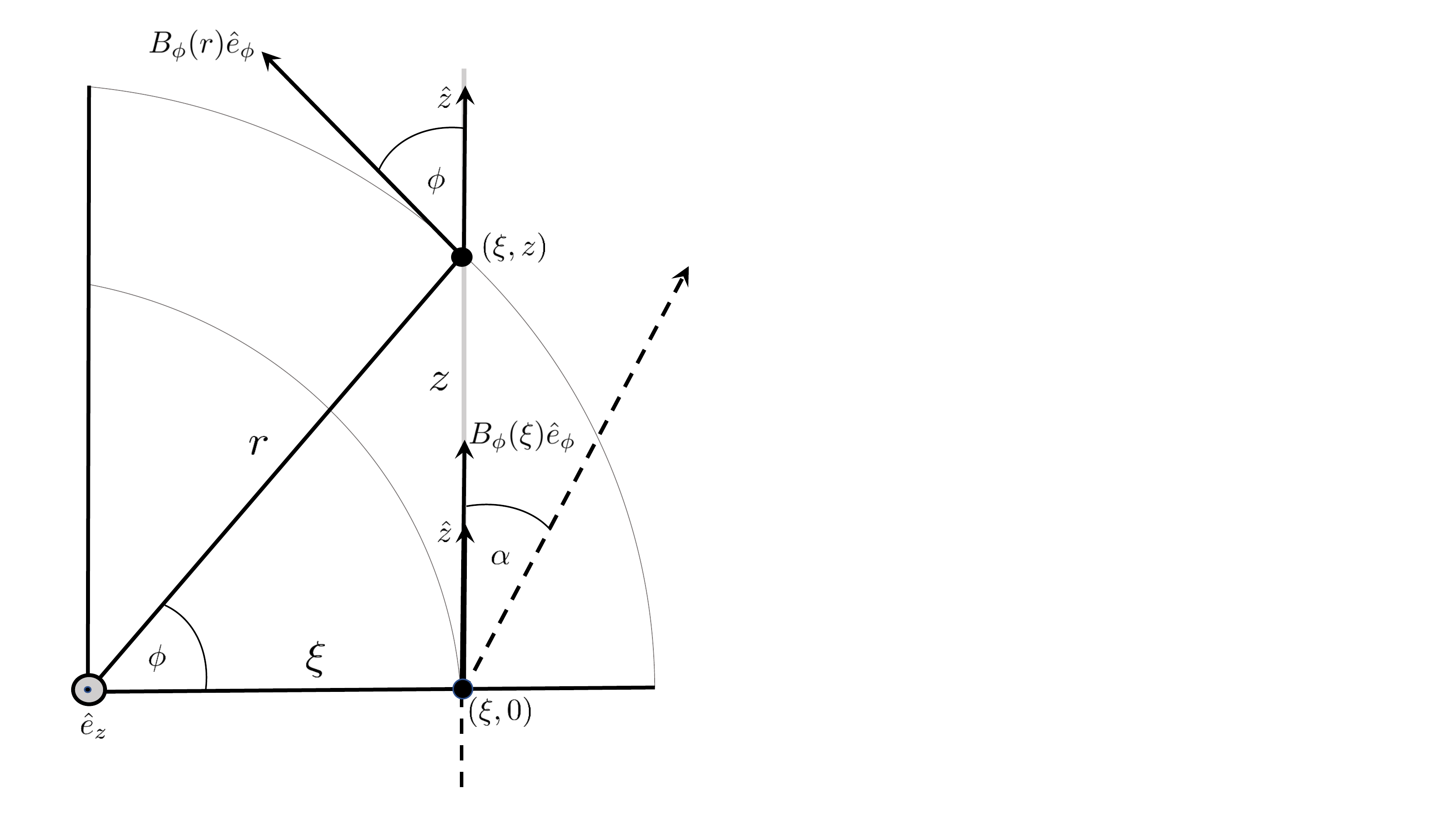}
\caption{An illustration of a ray (dashed line) being deflected by a magnetized plasma filament. The ray travels from source (bottom of figure) to observer (top of figure). The center of the filament is marked by the gray disk at the bottom-left of the figure. The horizontal black line represents the lens plane. The ray crosses the lens plane at the point $(\xi, 0)$, where $\xi=D_\text{d} \theta$ is the distance between the filament axis and the ray. The ray is deflected from its incoming direction, $\hat{z}$, by the angle $\alpha$. The path of the unperturbed ray is the vertical gray line. At $(\xi,0)$ the toroidal component of the magnetic field $B_\phi \hat{e}_\phi$ and the incoming ray $\hat{z}$ are co-linear, and point toward the observer along the ray in the $\hat{z}$ direction. At some farther point along the unperturbed trajectory, $(\xi, z)$, a distance $r$ from the origin, the toroidal field and the line of sight no longer align. This figure establishes that the projection of $B_\phi \hat{e}_\phi$ in the $\hat{z}$ direction depends on $\cos \phi=\xi/r$.}
\label{figGeom1}
\end{center}
\end{figure}

For filaments oriented perpendicular to the line of sight, only the toroidal component of helical field contributes to the RM, as this is the component that has a projection along the line of sight. Since the RM depends only on the projection of the field along the line of sight, we have
\be
\mathbf{B} \cdot d\mathbf{s} = \mathbf{B} \cdot \hat{z} dz = B_\phi \cos(\phi) dz = B_{||} dz,
\ee
where $\cos(\phi)=\xi/r$ is the opening angle between the magnetic field and the line of sight. The physical distance on the lens plane corresponding to the angle $\theta$ is $\xi=D_d \theta$, the distance of closest approach between an unperturbed ray and the lens center. We write an arbitrary distance along the unperturbed ray as $r=\sqrt{\xi^2+z^2}$. Let us once again refer to the geometry describing the lens and source planes in fig. \ref{figGeom2}. For a light ray that passes the filament, the axial field is always perpendicular to the light ray trajectory, so does not contribute. We only need to take into account the toroidal component of the field along the line of sight. Thus, in the absence of a radial magnetic field component, only the toroidal field contributes to the overall RM. We show these geometrical relationships graphically in fig. \ref{figGeom1}.

Having established the lensing properties of magnetized filaments, we now seek a physically motivated form for the heretofore undefined functions $n_\text{e}(r)$ and $B_\phi(r)$. To this end, we turn to magnetohydrodynamics to describe the physical characteristics of an ionized, magnetized filament.

\section{Magnetohydrodynamic Filament Models}
\label{sec:jason}

The introduction of a magnetic field provides an efficient mechanism for confining a plasma by trapping the charges to move along the field and providing resistance for particle motion across the field lines \citep{ISMReview}. In general, if the magnetic field of a filament has a toroidal component $B_\phi(r)$, the expression for the force on the filament becomes
\begin{equation}
    \mathbf{F}=-\hat{e}_r \frac{1}{8 \pi} \frac{1}{r^2} \frac{\partial}{\partial r} \left[ r^2 B_\phi^2(r) \right] ,
\end{equation}
which shows clearly that the force is directed radially inward, toward the center of the filament. This means that if the product $rB_\phi$ is an increasing function of r, the magnetic force on the filament matter points radially inward, acting to resist the pressure from the interior of the filament. Such a configuration with a toroidal magnetic field component thus provides a natural way to confine ionized filaments, and solves the over-pressure problem which plagues spherical, non-magnetized lens models. This magnetic field configuration is physically motivated, as filamentary structures occur throughout nature on a variety of scales and helical magnetic fields are closely related \citep{jf1, jf2}. 

To keep our intended meaning clear and to aid in interpretation, we have carefully labelled our variables with subscripts to stay organized. For a filament with radius $R$, the gas pressure within the filament is given as $P(r)$, with the pressure at the surface of the filament $P(R)=P_\text{S}$. Within the filament, the mean gas pressure is given as $<P>$. The ratio of these quantities is
\begin{equation}
\frac{P_\text{S}}{<P>} = 1-\frac{m}{m_\text{vir}} + \left( \frac{<P_\text{mag}>-P_\text{mag,S}}{<P>} \right)
\label{press}
\end{equation}
where $m$ is the line mass of the filament and the virial line mass is  
\be
m_\text{vir}=\frac{2\sigma^2}{G}. 
\ee
The average pressure due to the magnetic field is defined as
\begin{equation}
<P_\text{mag}> = \frac{1}{8 \pi V} \int B_\text{z}^2 dV,
\end{equation}
where only the field $B_\text{z}$ along the axis of the cylinder in the $\hat{e}_\text{z}$ direction contributes. The magnetic field pressure at the surface of the filament is
\begin{equation}
P_\text{mag,S} = \frac{B_\text{z,S}^2 + B_{\phi,\text{S}}^2}{8 \pi}.
\end{equation}
Since we are describing filaments on small scales in the ISM, we will assume the filaments are not self-gravitating, such that $m \ll m_\text{vir}$. With this assumption equation \ref{press} simplifies such that
\begin{equation}
<P>=P_\text{S} - <P_\text{mag}> + P_\text{mag,S}.
\label{simplePress}
\end{equation}
Thus, the average pressure inside the filament is greater than the surface pressure when $P_\text{mag,S}$ exceeds $<P_\text{mag}>$. This is easy to accomplish because $<P_\text{mag}>$ involves only $B_\text{z}$, whereas both the toroidal and poloidal components contribute to $P_\text{mag,S}$.

To progress from this point we must assume a particular form for the magnetic field of the filament. Suppose that the filament has a discontinuity in the axial field at the surface such that the coaxial field component $B_\text{z}$ within the filament is a multiple of the field at the surface and in the surrounding medium. This is due to the ISM being compressed to form the filament,
\begin{equation}
\left. B_\text{z} \right|_\text{inside} = a B_\text{z,S}
\end{equation}
with the amplification factor $a$ constant. Then, the average magnetic pressure depends on the pressure due to the $z$ component of the magnetic field (mag) on the surface ($S$) of the filament, where we have labelled our subscripts to carefully specify our intended meaning:
\begin{equation}
<P_\text{mag}> = \frac{a^2 B_\text{z,S}^2}{8 \pi} = a^2 P_\text{mag,z,S}.
\end{equation}
Now let us write the surface field pressure in terms of the $z$ and $\phi$ components, $P_\text{mag,S}=P_\text{mag,z,S}+P_\text{mag,$\phi$,S}$. Finally, combining all of this and factoring $P_\text{mag,z,S}$ gives
\begin{equation}
<P>=P_\text{S} + P_\text{mag,z,S} \left[ \frac{P_\text{mag,$\phi$,S}}{P_\text{mag,z,S}} - (a^2 -1) \right].
\end{equation}
In terms of the magnetic field components, we have
\begin{equation}
<P>=P_\text{S} + P_\text{mag,z,S} \left[ \frac{B_\text{$\phi$,S}^2}{B_\text{z,S}^2} - (a^2 -1) \right].
\end{equation}
Let us assume there are no axial surface currents flowing such that the filament has a toroidal field component $B_\phi$ which is generally continuous at the surface $S$ (Note that this configuration would have a toroidal surface current in the $\hat{e}_\phi$ direction due to the discontinuity in $B_\text{z}$). We define the pitch angle at the outer surface of the filament as
\begin{equation}
\tan \vartheta_\text{p} = \frac{B_\text{$\phi$,S}}{B_\text{z}} = \frac{B_\text{$\phi$,S}}{a B_\text{z,S}},
\end{equation}
such that
\begin{equation}
<P>=P_\text{S} + P_\text{mag,z,S} \left[ a^2\left( \tan^2 \vartheta_\text{p}-1 \right) +1 \right].
\label{eqp}
\end{equation}
Introducing the plasma $\beta$ parameter associated with the $z$ component of the field for the external plasma,
\begin{equation}
\beta_\text{Z} = \frac{P_\text{S}}{P_\text{mag,z,S}}
\end{equation}
and solving equation \ref{eqp} for the pitch angle, we find
\begin{equation}
\tan^2 \vartheta_\text{p} = \frac{\beta_\text{Z}}{a^2}\left( \frac{<P>}{P_\text{S}}-1 \right) + \left(\frac{a^2-1}{a^2} \right)
\label{eqp2}
\end{equation}
which gives us the winding angle between the coaxial and toroidal components as a function of average and surface pressure $<P>$ and $P_\text{S}$ respectively, the pressure ratio in the external plasma $\beta_\text{Z}$ and field strength inside the filament from the amplification factor $a$.

\begin{center}
\begin{figure}
\includegraphics[viewport=51 1 350 305, clip=true, scale=0.75]{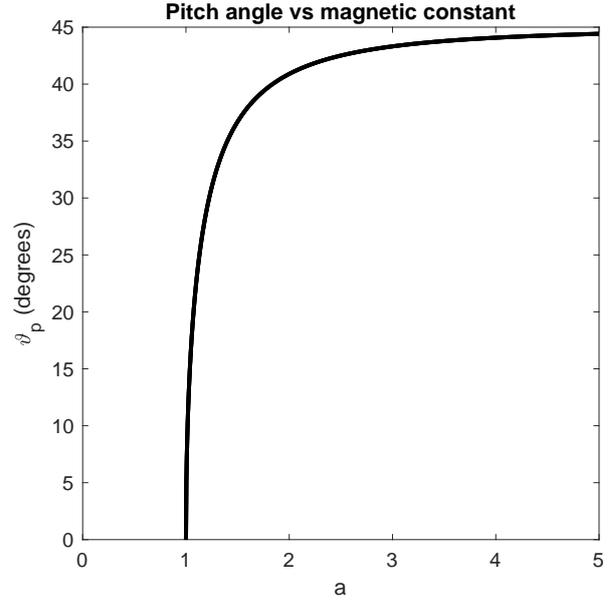}
\caption{Pitch angle $\vartheta_\text{p}$ vs magnetic constant $a$, in the limit $\beta_\text{Z} \rightarrow 0$. The maximum pitch angle in this limit is $\vartheta_p \rightarrow 45^\circ$.}
\label{figFil}
\end{figure}
\end{center}

There are two natural limits of this filament model to explore that will provide some insight as to the configuration of the magnetic field. First, let us consider a highly magnetized filament such that $\beta_\text{Z} \rightarrow 0$. Then only the second term in eq. \ref{eqp2} contributes,
\begin{equation}
\tan^2 \vartheta_\text{p} \approx \frac{a^2-1}{a^2}.
\label{eqpLim1}
\end{equation}
Since the square of the pitch angle tangent appears on the left hand side, real solutions exist only for $a>1$. As $a$ is increased significantly beyond $1$, the pitch angle approaches an upper limit of $45^\circ$. This is a particularly interesting result because it does not require a tightly wound field to confine the filament. We plot the pitch angle $\vartheta_\text{p}$ as a function of the magnetic coefficient $a$ in fig. \ref{figFil}. The second relevant limit is found by stating the pitch angle in terms of density, namely the average pressure $<P>=\rho \sigma_{v}^2$ where $\sigma_v$ is the internal filament velocity dispersion which may be due to thermal motion (ie, from the ideal gas law $\sigma_v^2=kT/[\mu m_\text{H}]$) or due to turbulence. The surface pressure can be written as $P_\text{S}=\rho_\text{S} \sigma_\text{ext}^2$ with $\rho_\text{S}$ the density on the surface of the filament (i.e., the ISM density) and $\sigma_\text{ext}$ is the velocity dispersion of the ISM external to the filament. Then equation \ref{eqp2} becomes
\begin{equation}
\tan^2 \vartheta_\text{p} = \frac{\beta_\text{Z}}{a^2}\left( \frac{\rho }{\rho_\text{S}}\frac{\sigma_v^2}{\sigma_\text{ext}^2}-1 \right) + \left(\frac{a^2-1}{a^2} \right).
\label{eqpLim2}
\end{equation}
Now let us write the conservation of mass during the compression of the ISM to form the filament. The density begins with the ISM density, $\rho_\text{S}$, which is identical to the surface density of the filament after compression. The initial cross-sectional area is $A_0$, such that conservation of mass relates the initial and final states
\begin{equation}
\rho A = \rho_\text{S} A_\text{0}
\end{equation}
and conservation of magnetic flux gives
\begin{equation}
B_\text{z} A = B_\text{z,S} A_\text{0}
\end{equation}
such that
\begin{equation}
\frac{\rho}{\rho_\text{S}} = \frac{B_\text{z}}{B_\text{z,S}} = a
\label{ratios}
\end{equation}
giving
\begin{equation}
\tan^2 \vartheta_\text{p} = \frac{\beta_\text{Z}}{a^2}\left( a\frac{\sigma_v^2}{\sigma_\text{ext}^2}-1 \right) + \left(\frac{a^2-1}{a^2} \right).
\label{eqpLim3}
\end{equation}
In the limit of large internal field $a \gg 1$,
\begin{equation}
\tan^2 \vartheta_\text{p}  \rightarrow \frac{\beta_\text{Z}}{a} \frac{\sigma_v^2}{\sigma_\text{ext}^2} + 1 \rightarrow 1,
\end{equation}
which again implies a pitch angle $\vartheta_p \rightarrow 45^\circ$. In the opposite limit, when the amplification factor $a=1$, there is no compression of the filament density or field and we find
\begin{equation}
\tan^2 \vartheta_\text{p}  = \beta_\text{Z} \left( \frac{\sigma_v^2}{\sigma_\text{ext}^2} - 1 \right).
\end{equation}
In this limit, the toroidal component $B_\phi$ exists only when the filament has a greater velocity dispersion inside than outside, $\sigma_v > \sigma_\text{ext}$. There is no need for confinement when the opposite is true, so the pitch angle $\vartheta_\text{p}$ has no real solution in that case. Recall that if $a=1$ the coaxial field $B_\text{z}$ and density $\rho$ are the same everywhere. In this case it is only the velocity dispersion that distinguishes the filament from its surrounding environment.

\subsection{Magnetic Filament Lifetimes}
\label{sec:mag2}

First, we refer to the discussion of filament lifespans in the review of \citet{ISMReview}. Let the neutral-ion timescale be $\tau_\text{ni}$, and the dynamical timescale be $\tau_\text{dyn}$. We will also take $v_\text{d}$ to be the drift velocity and $c_\text{s}$ the sound speed in the filament. Summarizing the results presented by \citet{ISMReview} we have,
\begin{equation}
\frac{c_\text{s}}{v_\text{d}}=c_\text{s} \frac{\tau_\text{ni}}{R} = \frac{\tau_\text{ni}}{\tau_\text{dyn}},
\label{vRatio}
\end{equation}
where we have corrected the typo that showed the ratio on the left hand side erroneously inverted. However, the first equality in the expression above is problematic. In the middle term, \citet{ISMReview} make the association between the drift velocity, radius of the filament and neutral-ion timescale. This is not generally true as the drift velocity depends on the mean free path $\ell$, rather than the filament radius $R$ as in eq. \ref{vRatio}:
\begin{equation}
v_\text{d}=\frac{\ell}{\tau_\text{ni}}.
\label{eq:vd}
\end{equation}
The difference here is that the mean free path can be much smaller than the radius of the filament, $\ell \ll R$. \citet{ISMReview} argue that $c_\text{s}/v_\text{d} =1$ using $R \approx 35$ AU implies a minimum filament scale at $R\approx 35$ AU. But if $\ell$ is used in the calculation instead, $c_\text{s}/v_\text{d} =1$ can be found for much smaller values using $\ell \ll R$, which argues against the minimum scale limit for magnetically confined filaments discussed in \citet{ISMReview}. Let us write $R=y \ell$ where $y=R/ \ell \gg 1$ such that
\begin{equation}
v_\text{d}=\frac{R}{y \tau_\text{ni}}.
\end{equation}
Thus, equation \ref{vRatio} becomes
\begin{equation}
\frac{c_\text{s}}{v_\text{d}} = y \frac{\tau_\text{ni}}{\tau_\text{dyn}}.
\end{equation}
 Stability against dissipation of the field by neutral-ion drift requires $\tau_\text{dyn} \gg \tau_\text{ni}$, such that
\begin{equation}
 \frac{\tau_\text{ni}}{\tau_\text{dyn}}= \frac{1}{y} \frac{c_\text{s}}{v_\text{d}} \ll 1
\end{equation}
which is guaranteed even for $v_\text{d} \approx c_\text{s}$ if $y \gg 1$. From this argument, AU-scale filaments in the ISM are unstable only when the condition $y \gg 1$ is not true. This occurs when the radius of the filament $R$ is comparable to the neutral-ion mean free path $\ell$, which is the limit considered in \citet{ISMReview}, who estimated the minimum radius for filament stability to be $\approx 35$ AU. Generally, smaller filaments can be much longer lived than suggested by this limit provided their radii exceed the mean free path. Assuming a cross section $\sigma=\pi a_0^2$ where $a_0$ is the Bohr radius, and a density typical of ESEs, $n_S=3000$ cm$^{-3}$ (ie, a factor of $10^4$ higher than the warm ionized ISM), we find the mean free path $\ell=0.25$ AU. It is interesting that this is the scale usually associated with ESEs. Since the mean free path can be smaller if higher densities are considered, extremely dense filaments may be stable down to even smaller radii. However, the filament stability condition requires $\ell \ll R$. At some point, filaments with radii near the mean free path length $\ell \approx R$ should also be near the edge of stability. This appears to be required to accommodate the smallest size of ESE lenses, on the order of $0.1$ AU.

Let us try to analytically estimate the lifetime of a magnetically confined filament in more detail. We follow the approach of \citet{jf1} and \citet{jf2} to examine the conditions for stability for a partially ionized filament relevant for ESE modeling. Filaments are stable provided that their fragmentation time is longer than the dynamical timescale of the filament. In \citet{jf1}, the authors define the squared growth rate $-\omega^2$ of the most significant unstable mode which dominates fragmentation of the filament. While the squared growth rate is shown with a negative sign, it is found through an eigenvalue procedure where $-\omega^2$ may be purely real or imaginary. A perturbation of the cloud oscillates about the equilibrium state and is stable when $\omega$ is real. When $\omega$ is imaginary, the system is unstable and the perturbation grows exponentially, leading to the fragmentation of the filament. We are interested in the maximum growth rate of these destructive unstable modes as they will limit the life of the filament. For simplicity, we neglect the sign of the squared growth rate, but this does not affect the results of our calculation.

\begin{center}
\begin{figure}
\includegraphics[viewport=30 1 381 306, clip=true, scale=0.68]{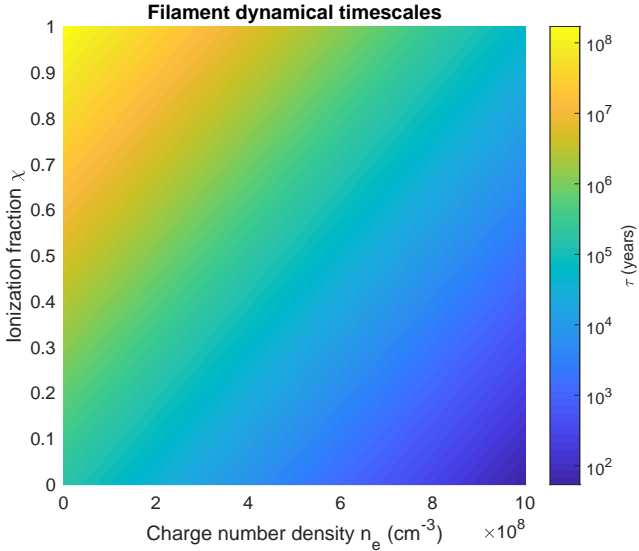}
\caption{Estimates for the lifetime $\tau$ of a magnetically confined filament with toroidal magnetic field. As the ionization fraction $\chi$ is increased, the corresponding age of the filament increases. The lifetimes reach from a few hundred years for low $\chi=10^{-6}$ to a the order of a hundred million years for $\chi=1$. For all ionization fractions these life times are in excess of the average ESE duration of weeks to months. Therefore, magnetic confinement of filaments is an extremely effective mechanism for confining filamentary structures on the scales relevant for the phenomena.}
\label{fig:timescale}
\end{figure}
\end{center}

We define the dynamical time scale, which depends on the density of the filament,
\begin{equation}
\tau_\text{dyn}= \frac{1}{\sqrt{4 \pi G \rho}}
\end{equation}
using $\rho = \mu m_H n$ where $\mu$ is the mean molecular mass, $m_H$ the mass of the hydrogen atom and $n$ the central number density of the filament. Using the ionization fraction $\chi$, we can put this in terms of the charge number density $n_e=\chi n$. Putting all of this together gives the dynamic timescale,
\begin{equation}
\tau_\text{dyn}= \sqrt{\frac{\chi}{4 \pi G \mu m_H n_e}}.
\end{equation}
Let us call $x$ the maximum of the normalized squared growth rate for magnetized filaments with helical fields,
\begin{equation}
x=\frac{\omega^2}{4 \pi G \rho}.
\label{xeq}
\end{equation}
The inverse growth rate gives the growth timescale,
\begin{equation}
\tau=\frac{2 \pi}{|\omega|}
\end{equation}
which, along with eq. \ref{xeq}, gives
\begin{equation}
\tau= \tau_\text{dyn}\frac{2 \pi }{\sqrt{x}}.
\end{equation}
We estimate $x$ using the numerical data presented in fig. 9 and table 1 of \citet{jf2} for a filament with a helical magnetic field. From this study, the maximum $x$ for a weakly self-gravitating filament is found to be on the order of $0.01$, which we adopt for our analysis. The ionization fraction $\chi$ is the remaining unknown quantity, which we vary from $10^{-6}$ to $1$. Generally, ionization fractions in the range of $10^{-6}$ to $10^{-4}$ are appropriate for molecular gas in giant cloud complexes. Since we are dealing with atomic gas in the ISM, we expect the ionization fraction to be higher than for molecular gas. Since $\tau_\text{dyn}$ is proportional to $\sqrt{\chi}$, the most conservative age estimate of filaments should occur for small $\chi$. Even using the unrealistically small ionization fraction of $\chi=10^{-6}$, filaments with the density usually associated with ESEs ($n_\text{e}=10^6$ cm$^{-3}$) can be stable on the order of a thousand years. As $\chi$ is increased to more reasonable values, the timescales correspondingly grow. The relationship between filament lifetime $\tau$, ionization fraction $\chi$ and charge number density $n_\text{e}$ is shown in fig. \ref{fig:timescale}. This estimate suggests that magnetically confined filaments with toroidal fields can be remarkably stable and long-lived assuming physical conditions relevant for ESEs.

While our filament models are non self-gravitating, some of the filaments considered by \citet{jf1, jf2} were pressure-truncated and weakly self-gravitating. Therefore, the stability calculations from that study are relevant for our models as well. The dominant mode considered in \citet{jf2} is the fundamental ``sausage'' mode, and is active for all toroidal magnetic fields, even when the field is very weak \citep[i.e., when $B_\phi \rightarrow 0$;][]{jf1,jf2}. The low-mass regime, where we consider filaments with $m/m_\text{vir} \ll 1$, should be even more stable than the weakly self-gravitating cases. A future study on the effect of higher modes is also planned.

\section{Example Force-Free Lenses}
\label{sec:models}

With an MHD model of a magnetized, ionized filament in hand, we turn to determining the optical properties of such an object acting as a plasma lens. We begin with the simplest possibility, a non-magnetized filament which depends only on the contribution from the DM term. Next, we build on this simple model by adding the magnetic field and including the deflection contribution from the RM term. We also evaluate the observational prospects for detecting the magnetic field within a filament lens.

\subsection{A Non-Magnetized Filament with Discontinous Density}
\label{sub:nonMagFilModel}

The MHD filament developed in Section \ref{sec:jason} is unusual as a lens model in that it contains constant density, and the surface of the filament is discontinuous in both density and poloidal field (shown in eq \ref{ratios}). The number density outside the filament is $n_{S}$, the density of the ambient ISM. Inside the filament the density increases to $n=a n_S$ with amplification factor $a>1$. Note that the density both inside and outside the filament is constant. However, the filament is assumed to have a circular cross-section with radius $R$. Therefore, the projected density of the filament varies over the lens plane, despite the constant density within the lens. Physically, a light ray through the center of the filament passes through more of the high density plasma than a ray that traverses the lens near its edge due to the curved surface of the lens. Thus, a cylindrical, constant density filament does produce a DM gradient over the plane of the sky due to the cylindrical geometry. Ray trajectories that do not intersect the filament remain undeflected since the integral along these rays is constant.

Due to the model discontinuities, we split the integral along the ray trajectory into three parts. The first leg of the journey integrates the undeflected ray path from the source which we approximate by integrating from a great distance $-\delta$ to the surface of the filament at $-z_\text{p}$ and passes through a uniform medium of density $n_\text{S}$. Suppose the undeflected ray trajectory passes through the filament interior with uniform density $n=a n_\text{S}$ and exits the filament at a point $+z_\text{p}$. For a filament with radius $R$, the ingress and egress points are given by $z_\text{p}=\sqrt{R^2-\xi^2}$, with $\xi=D_\text{d} \theta$ as the closest approach between the ray and the center of the filament. Finally, after exiting the filament, the ray travels from the surface at $+z_\text{p}$ to the observer a great distance away at $+\delta$. Thus, we can project the lens quantities along the line of sight in a piecewise sum. In this way, we find
\be
\begin{array}{ll}
DM = \int_{-\delta}^{+\delta} n(r) dz & \\
= n_\text{S} \left(\int_{-\delta}^{-\sqrt{R^2-\xi^2}} dz + a \int_{-\sqrt{R^2-\xi^2}}^{+\sqrt{R^2-\xi^2}}dz + \int_{+\sqrt{R^2-\xi^2}}^{+\delta} dz \right), &
\label{deflDisc}
\end{array}
\ee
and with the filament radius $R=D_\text{d}\theta_\text{R}$ gives
\be
DM(\theta)=2 n_\text{S} (a-1)D_\text{d}\theta_R\left( 1-\frac{\theta^2}{\theta_R^2} \right)^\frac{1}{2} +2n_\text{S} \delta
\ee
for rays with $|\theta|<\theta_R$, which pass through the lens. Since the limits of the integral depend on $\xi$ (and therefore by extension $\theta$), the gradient $\nabla_\theta$ in the calculation of the deflection angle cannot be freely moved inside the integral to operate on the integrand before the integration is complete. Rays with $|\theta|\geq\theta_\text{R}$ are exterior to the lens, and this means the integral in eq. \ref{deflDisc} can be done without the need to split it into parts. Since the ISM is constant density outside the lens, the DM is constant and the deflection angle $\alpha_{DM}$ vanishes. Thus, rays that do not pass through the lens do not experience any deflection.

\begin{figure}
\includegraphics[viewport=50 0 355 306, clip=true, scale=0.80]{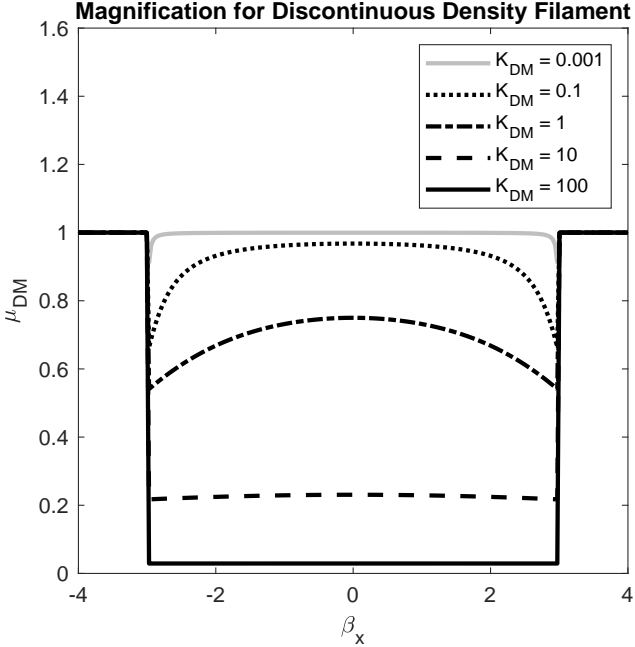}
\caption{Magnification for the filament lens with discontinuous density. The parameters chosen here are $a=2$ and $\theta_R=3$. The $a$ parameter modifies how rapidly the magnification changes with frequency due to the lens overdensity, and $\theta_R$ determines the width of the exclusion region formed.}
\label{fig:novDM}
\end{figure}

In the DM expression, we assume a great distance between lens, source and observer. We integrate from $-\delta$ to $\delta$, where $\delta$ is very large. In the Born approximation, we calculate the integral along an undeflected ray from $-\infty$ to $+\infty$. In this limit the DM (lens potential) becomes formally infinite due to the term containing $\delta$. This arises from the assumption of a medium of constant density $n_S$ that fills all space outside the filament. Despite the appearance of this large term, the deflection angle depends on the gradient of the lens potential. Therefore, since $\delta$ is large but constant, depending only on the limits of the integration, it does not contribute to the gradient. Thus, the deflection angle itself remains finite. The first term in the expression above is indeed a changing component across the projected filament. Thus, despite the complications from the discontinuities, the deflection angle in this case turns out to be surprisingly simple
\be
\alpha_\text{DM}(\theta)=- K_\text{DM} \frac{\theta}{\theta_R} \frac{2(a-1)}{\left( 1-\frac{\theta^2}{\theta_\text{R}^2} \right)^\frac{1}{2}}
\label{deflDM}
\ee
with the constant from the definition of the effective lens potential, eq.
\ref{potDefn},
\be
K_\text{DM}=\frac{\lambda^2}{2 \pi} \frac{D_\text{ds}}{D_\text{s} } r_\text{e} n_\text{S}.
\label{defKDM}
\ee
The inverse magnification for the constant density filament is
\be
\mu_\text{DM}^{-1} = 1-\frac{d \alpha_{DM}}{d \theta} = 1+\frac{K_\text{DM}}{\theta_R}\frac{2(a-1)}{\left( 1-\frac{\theta^2}{\theta_\text{R}^2} \right)^\frac{3}{2}} ,
\label{muDM}
\ee
with $|\theta|<\theta_R$. An example of lensing by the non-magnetized filament is shown in fig. \ref{fig:novDM} as a function of frequency. The effect of this lens is very simple, as it essentially just defines an ``exclusion region'' which forms extremely de-magnified images due to the diverging effect of plasma lensing. Since the lens is discontinuous, eqs. \ref{deflDM} and \ref{muDM} are strictly defined only inside the lens $|\theta|<\theta_R$. At the exact boundary $|\theta|=\theta_R$, no deflection occurs. The thin lens equation says that $\beta=\theta=\theta_R$ here. The light curve outside the lens is unity everywhere since no refraction occurs there. 

\subsection{A force-free helical field with a toroidal power-law component}
\label{sec:magFilModel}

To evaluate the RM, we must assume a specific form for the toroidal magnetic field. Suppose the toroidal field has a simple power-law form,
\begin{equation}
B_\phi(r) = B_0 \frac{R}{r},
\label{magParam}
\end{equation}
with the constant $B_0$ as the strength of the field at the filament surface $r=R$. This is a force-free magnetic field, since $rB_\phi$ is a constant, so it does not provide any radial force. We choose this form as a convenient analytical example, but we expect real magnetic fields to decrease more slowly than a $r^{-1}$ dependence. Nevertheless, the power-law field provides a clear and intuitive example of lensing by a magnetized filament.    

With this toroidal field, we calculate the deflection angle due to the RM from eq. \ref{RMdefn}, which we split up to address the discontinuity in the lens model analogously with eq. \ref{deflDisc}. Putting this approach together with the assumed field $B_\phi(r)$, using $\cos \phi = \xi/r$, $r=\sqrt{\xi^2+z^2}$ and the discontinuous lens density profile, we find the RM which gives
\be
\psi_{RM}=K_{RM}\left( 2(a-1) \theta_R \tan^{-1}\left(\frac{\sqrt{\theta_R^2-\theta^2}}{\theta} \right) + \text{sign}(\theta) \pi \right)
\ee
where we have defined 
\be
K_{RM}= \lambda^3 \frac{D_{ds}}{D_s} \frac{r_\text{e}^2}{4 \pi^2 e}B_0 n_S
\label{defKRM}
\ee
and sign($\theta$) changes from one side of the lens to the other. However, on both sides this quantity is constant, so it does not impact the calculation of the gradient. The RM is shown in fig. \ref{fig:novRM} as the thick gray lines. We write the argument of the inverse tangent carefully to keep the denominator separate from the numerator. The sign of the denominator is important since it is linear and causes the inverse tangent function to flip signs on opposite sides of the filament. When the magnetic field points toward the observer, $B_{||}(r)>0$ and the lens potential (RM) is positive, and vice-versa when the magnetic field points away from the observer. From the potential, we arrive at the deflection angle
\be
\alpha_{RM}(\theta)=-\frac{2(a-1)K_\text{RM}}{ \left( 1-\frac{\theta^2}{\theta_R^2} \right)^{\frac{1}{2}} }
\ee
Note that the deflection angle is negative for all $\theta$, as we expect from examination of fig. \ref{fig:novRM}. Reversing the parity of the lens potential (ie, reflecting the effective potential about the origin) maintains the sign of the derivative. Note that this lens does not produce any caustics or critical curves.

Now, let us investigate the scale of both deflection angles, as given by their multiplicative coefficients $K_{DM}$ and $K_{RM}$, in equations \ref{defKDM} and \ref{defKRM}, respectively. To compare the strength of refractive deflection due to DM and RM terms, we write the ratio of the coefficients as
\be
\frac{K_{RM}}{K_{DM}}=\frac{c}{2 \pi \nu} \frac{r_e}{e} B_0 \approx 10^{-9}
\ee
where we have written $\lambda=c/\nu$ in GHz and the average galactic field as $B_0 \approx 1 \mu G$ \citep{cff, RMconstraints, ISMReview}. In fact, this holds true for all practical observing bands. Even for extremely low frequency radiation on the order of $10$ MHz, the limit of radio-wave transmission through the Earths atmosphere, the deflection due to RM is still $6$ orders of magnitude smaller than the DM term. This means that the deflection and magnification due to the RM term are negligible compared to the contribution from the electron density.

Despite the apparent lack of observable effects on the magnification, there are still signatures that could suggest a toroidal magnetic field component is present. First, the lens RM (and by extension the effective lens potential for magnetized plasma) is a measurable quantity. We predict the form of the lens RM, shown in fig. \ref{fig:novRM}, which suggests a magnetic field pointing away from the observer on one side of the lens and toward on the other. If the RM of a lens could be repeatedly measured while an ESE was occurring, it would provide strong support in favor of or against a magnetized filament lens perpendicular to the line of sight. Second, \citet{filamentRM} have shown that the difference between the deflection angles $\alpha_{DM} \pm \alpha_{RM}$ of the two polarization states may produce a differential time delay that is in principle observable. However, this effect is practical to measure only for the the large RMs observed along the line of sight to the Galactic centre magnetar, J1745-2900, and the repeating fast radio burst source FRB121102. \citet{filamentRM} also used a filament lens with a constant magnetic field in the direction of the observer. Due to the small RM of the lenses discussed here, we believe this delay would be too small to measure for our model filaments.
 
\subsection{A broken power-law model}
\label{sub:broken}
 
\begin{figure}
\includegraphics[viewport=46 0 355 306, clip=true, scale=0.80]{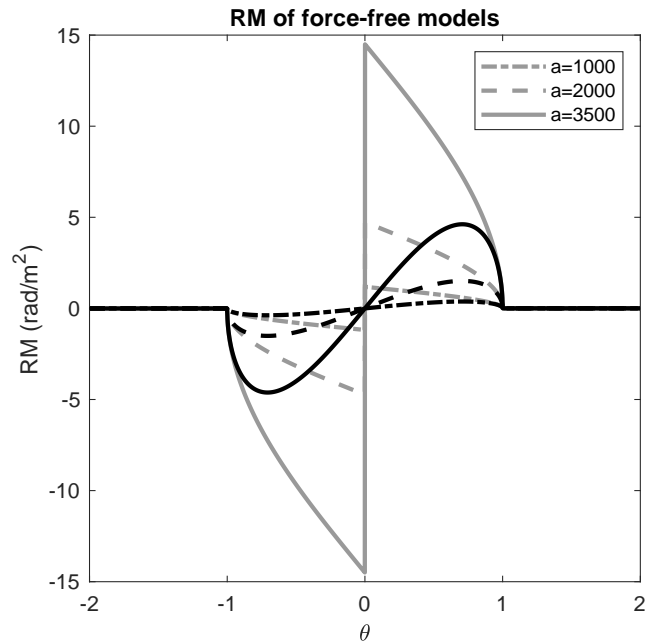}
\caption{RM of the piecewise (black) and broken power-law (gray) toroidal magnetic field for a variety of amplification factors $a$. We have chosen the surface fields to be the average galactic field $B_{z,S}=3.2$ $\mu$G, and ISM density $n_S=0.3$ cm$^{-3}$. The radius of the filament is assumed to be $R=0.1$ AU. At these amplification factors, relevant for ESEs, the poloidal and toroidal fields are equal in scale.}
\label{fig:novRM}
\end{figure}

Since the RM provides a valuable and potentially observational signature of the magnetic field of the lens, let us consider a second model, this time in terms of a toroidal field that is given as a piecewise broken power-law. The force-free $r^{-1}$ power-law field considered previously is the field expected from a thin current carrying wire at the center of the filament. A more realistic field should change its behaviour from the interior to the exterior of the filament. Inside we want the field to satisfy the more realistic boundary condition $B_\phi(0)=0$, growing monotonically as $r$ increases from  from the center of the filament to its surface at $r=R$. Exterior to the filament, the field should drop off with distance from the filament surface following a $r^{-1}$ power-law, assuming that the $B_\phi$ producing current is confined to the filament. Therefore, let us assume a piecewise toroidal power-law, such that
\be
B_\phi(r)=\left\{
\begin{array}{ll}
B_0 \frac{r}{R} & \text{Interior, } 0 \leq r \leq R \\
B_0 \frac{R}{r} & \text{Exterior, } r > R
\end{array}
\right.
\ee 
This will not excessively complicate the calculation since we are splitting up the integral in a piecewise fashion for the filament density anyway. Therefore, following along with the previous calculation we find, for $\theta<\theta_R$, 
\be
\begin{array}{ll}
\psi_{RM}= & K_{RM}\theta_R\left[ 2a\frac{\theta}{\theta_R}\left(1-\frac{\theta^2}{\theta_R^2} \right)^\frac{1}{2} \right. \\
 & \left. - 2 \tan^{-1}\left( \frac{\sqrt{\theta_R^2-\theta^2}}{\theta} \right)+\text{sign}(\theta)\pi \right].
\end{array}
\ee
Outside the filament, the effective lens potential is simply $\psi_{RM}=\text{sign}(\theta) K_{RM} \theta_R \pi$, a constant. With the effective potential in hand, all other quantities relevant for lensing may be found through derivatives. The RM is shown in fig. \ref{fig:novRM} as the thin, black lines. This field is more realistic in the sense that it accounts for the finite physical size of the filament, however we point out that it is also a force-free field and thus not sufficient to constrain realistic filaments. 

Despite the simple models studied here, the RM values they predict are interesting. Consider the RM plots shown in Fig. \ref{fig:novRM}. The figure shows the broken power-law toroidal field model with an amplification factor $a=3500$ and the piecewise field model with $a=2000$ varying between roughly $-5<RM<5$ rad m$^{-2}$, or $\Delta RM\approx10$ rad m$^{-2}$. In fact, the RM of the extragalactic radio source 1741-038 was measured while it was undergoing an ESE in 1992 using the VLA \citep{cff}. No change was detected to within $\Delta RM \leq 10$ rad m$^{-2}$. Similar RM constraints were obtained by \citet{RMconstraints} examining the long-term behaviour of the variable quasar J1819+3845. A constraint of $\Delta RM \leq 10$ rad m$^{-2}$ was also obtained for this source. It is interesting that our filament models can easily accommodate such small changes in RM. For filaments with broken power-law fields and radii on the order of $R=0.1$ AU, the filament amplification factor could be up to maximum $a=3500$ and produce changes in RM consistent with the constraint set by these observations. Thus, though the filaments may be quite dense and have high toroidal magnetic fields at their surfaces, the radius of the filaments are very small, causing the RM change to be smaller. 

\section{Discussion and Conclusions}
\label{sec:discussionConclusions}

Filament models help to relieve the problem of high internal pressures of plasma lenses required to produce ESEs. In this work, we have shown that it is possible to produce cylinder lenses that reproduce the magnification of spherical lenses when viewed perpendicular to the lens axis. However, the resulting DM of the lens models results in an unphysical scenario with a cylinder lens that is surrounded by a converging under-dense medium. In another example, the lens charge density was found to increase with distance from the filament. These examples are shown in fig. \ref{fig:lensRecon}. This behaviour does not represent a physically-realizable scenario, so we conclude that realistic filament lenses perpendicular to the line of sight should produce distinct magnification of background sources compared to spherical lenses, unless viewed in a different orientation (such as end-on). This illustrates one key difference between plasma lenses and gravitational lenses. 

The second part of our work provides a further investigation of filametary magnetic fields. After demonstrating the spherical-cylindrical lens degeneracy, we took a census of the magnetized plasma filaments from the literature and introduced a filament with a helical magnetic field. If filament lenses in the ISM are confined by $B_\phi$ dominated helical magnetic fields, the lens pressure problem may be solved by the inward radial force acting to support the over-pressured filament. In this picture, filaments may act as lenses in any orientation with respect to the observer. Filaments that are nearly aligned with the line of sight appear to produce the most high-density events, while more gently inclined filaments appear to have lower density and should show a toroidal magnetic field component. 

Filaments with helical magnetic fields occur over many physical scales in nature, appearing in environments that range from from cold molecular cloud complexes to the jets of quasars. We developed a novel MHD model that describes a non self-gravitating ionized filament that can act as a plasma lens. We have shown that if filaments on AU scales exist in the ISM with mean free path $\ell \ll R$, helical magnetic fields are an exceptionally effective mechanism for confining them. In addition, we studied a range of ionization fractions that produce filaments with lifespans of thousands of years for the range of charge densities necessary for ESEs.

In Section \ref{sec:mag2} we calculated the lower-limit for the stability of ionized filaments should be $\ell=0.25$ AU at $10^4$ times the ambient ISM density, which is a density associated with ESEs. Even on these scales, the filaments we discuss in this work are much larger than the filaments suggested by \citet{noodles2} in their noodle model for scintillation arcs. It may be that both types of structures have their own particular regimes of applicability and mechanism for formation, namely that large stable filaments may provide the long timescales needed in ESEs while the scintillation of radio sources may occur due to smaller, dynamically unstable structures.

The nature of the magnetic field can be observationally probed with measurements of the lens RM. Our filament with a toroidal field component will produce a relatively small RM that varies perpendicular to the axis of the filament. Measurement of such a variation in RM would provide evidence for the presence of a toroidal magnetic field component. However, we have shown that in practice the RM produced by the filaments may be quite small due to the assumed radius of the filament, $0.1$ AU which the RM is linearly proportional to. The small radius can hide the large fields within the surface of the filament. We argue that such a configuration can easily fall within the observational constraints for a broken power-law model up to amplification factor $a=3500$ while avoiding observational detection through changes in RM on the order of or below $10$ rad m$^{-2}$. Projection changes the apparent properties of a cylinder lens and the parallel component of the tilted filament magnetic field will include contributions from both components, $B_{||} \rightarrow B_\phi \sin i + B_z \cos i$. If $i=\pi/2$, the filament is perpendicular to the line of sight and we recover the results discussed in this work.

The difficulty with magnetically confined filaments on the AU-scales required for ESEs is the mechanism of their creation. In large scale molecular filaments, the inward gravitational force provides an additional force acting on filamentary structures. However, our model filament is non self-gravitating since the line mass of these AU-scale ionized filaments is much smaller than the line mass of molecular filaments. The problem of how a helical field can form on these scales is left as an open question. However, our model shows that non self-gravitating filaments are consistent with MHD and can be stable for large radii $>1$ AU. Their formation may require a turbulent medium to wind up the magnetic fields that contain the filaments, or some other mechanism to generate torsional Alfven waves, such as has been proposed for the molecular filaments near the Galactic center, known as molecular tornadoes \citep{auFiege}.

Our model filament provides an analytical solution for a simple case (the force-free toy model developed in Section \ref{sec:models}). We choose to expand on this case for analytical clarity. In fact, beyond the $1/r$ limiting case that we investigated, a more general choice for the toroidal magnetic field of the filament is
\be
B_\phi(r)=\left\{
\begin{array}{ll}
B_0 \left( \frac{r}{R} \right)^{h_1} & \text{Interior, } 0 \leq r \leq R \\
B_0 \left( \frac{R}{r} \right)^{h_2} & \text{Exterior, } r > R
\end{array}
\right.
\label{magParam2}
\ee 
with the constant $B_0$ as the strength of the field at the surface of the filament $r=R$, and $h_1>0$. External to the filament the product $r B_\phi = B_0 R^{h_2} r^{ 1-h_2 }$ is an increasing function provided $0<h_2<1$. 
With this condition a simple power-law magnetic field provides radial force to constrain an ionized filament while simultaneously satisfying the $B_\phi(0)=0$ boundary condition. The $h_1=h_2=1$ extreme is the force-free limit considered in our work. In fact, the virial equation (i.e., eqs. \ref{press} and \ref{simplePress}) used elsewhere in this work and derived in \citet{jf1} does not apply to the case in which $B_\phi \propto 1/r$ and generally requires a faster decrease to be valid. While it is possible to produce solutions based on the more realistic power-law field in eq. \ref{magParam2}, difficulties arise due to the surface discontinuity in the lens density. Splitting up the integral in the required way produces complicated solutions in terms of hypergeometric functions. Since the magnification depends on the second derivative of the potential these expressions become extremely complicated very quickly. Due to these complicated analytical forms we choose not to consider these solutions as instructional examples for illustrating the lensing behaviour of magnetized filaments. A thorough and full investigation of realistic magnetized filaments requires more involved numerical analysis that is beyond the scope of this work. Instead, we intend to present an example that is simple and clear to illustrate the physical effect of a toroidal field component on the lensing properties of a filament. Thus, we will use the less realistic but analytically tractable $h_1=h_2=1$ limiting case and explore more realistic plasma lens filaments with $h_1>0$, $h_2<1$ in future work.

Beyond the details of the magnetic field, the filament lenses developed in this work are toy models. Since they are derived using the virial theorem, we can think of the filaments with constant density interiors as an ``average'' filament profile. In fact, it is the constant density interior that causes difficulties for these filaments as a lens model. The model features large discontinuities in the density and polar magnetic field at the surface. Due to the piecewise nature of the lens, we find it does not produce any caustics, and so cannot describe realistic ESEs. For more realistic models we require equilibrium solutions, which will feature more realistic continuous density distributions, smooth gradients, and lens substructure, which will provide more varied and rich lensing behaviour.

Finally, we emphasize once again that while we have discussed filamentary lens models in this work, a charge density that projects to a strip in the lens plane can also be explained through an inclined sheet of charge. Such projected charge densities could equally-well describe either case. The sheet model has been explored at length in the literature \citep{spSheets, ESE2}. A projected strip on the lens plane is also the special case of an elliptical plasma lens in the limit that the axis ratio vanishes $q \rightarrow 0$ \citep{erRogers2}.

\section{Acknowledgements}
We would like to thank the anonymous referee for thoughtful suggestions that improved the clarity and flow of our manuscript. We also thank Kelvin Au for many interesting discussions and for proofreading the manuscript. A.R., A.M. and B.P. wish to acknowledge the contribution of the Brandon University Research Committee and the Canada Summer Jobs Program. X.E. is supported by NSFC Grant No. 11473032 and No. 11873006.

\end{document}